\documentclass[draftclsnofoot,onecolumn]{IEEEtran}

\ifCLASSINFOpdf
\else
\fi

\newtheorem{theorem}{Theorem}[section]
\newtheorem{definition}[theorem]{Definition}
\newtheorem{lemma}[theorem]{Lemma}
\newtheorem{corollary}[theorem]{Corollary}

\newtheorem{exmp}[theorem]{Example}

\newcommand{\EXAMPLE}{\begin{exmp} }
\newcommand{\eoEXAMPLE}{\end{exmp}}

\newcommand{\ignore}[1]{}
\newcommand{\tabincell}[2]{\begin{tabular}{@{}#1@{}}#2 \end{tabular}}
\usepackage{amsmath}
\usepackage{amssymb}
\usepackage{slashbox}
\usepackage[square, comma, sort&compress, numbers]{natbib}

\begin{document}
%
\title{New families of Strictly optimal Frequency hopping sequence sets}
%
%
%

\author{Jingjun Bao
\thanks{
}
\thanks{ J. Bao is with the Department of Mathematics, Soochow University, Suzhou 215006, P. R. China. E-mail:  baojingjun@hotmail.com.
}
}

%
%

\markboth{}%
{Shell \MakeLowercase{\textit{et al.}}: Bare Demo of IEEEtran.cls for Journals}
%



\maketitle



%
\IEEEpeerreviewmaketitle
\begin{abstract}
Frequency hopping sequences (FHSs) with favorable partial Hamming correlation properties have important applications in many synchronization and  multiple-access systems. In this paper, we investigate constructions of FHS sets with optimal partial Hamming correlation. We present several direct constructions for balanced nested cyclic difference packings (BNCDPs) and balanced nested cyclic relative difference packings (BNCRDPs) such that both of them have a special property by using trace functions and discrete logarithm. We also show three recursive constructions for FHS sets with partial Hamming correlation, which are based on cyclic difference matrices and discrete logarithm. Combing these BNCDPs, BNCRDPs and three recursive constructions, we obtain infinitely many new strictly optimal FHS sets with respect to the Peng-Fan bounds.
\end{abstract}


\section{Introduction}

\IEEEPARstart{F}{requency} hopping (FH) multiple-access is widely used in the modern communication systems such as ultrawideband (UWB), military communications, Bluetooth and so on, for example, \cite{B2003}, \cite{FD1996},
\cite{YG2004}. In FH  multiple-access communication systems, frequency hopping sequences are employed to specify the frequency on which each sender transmits a message at any given time. An important component of FH spread-spectrum systems is a family of sequences having good correlation properties for sequence length over suitable number of available frequencies. The optimality of correlation properties is usually measured according to the well-known Lempel-Greenberger bound \cite{LG1974} and Peng-Fan bounds \cite{PF2004}. During these decades, many algebraic or combinatorial constructions for FHSs or FHS sets meeting these bounds have been proposed, see \cite{BJ},  \cite{CJ2005}-\cite{CY2010}, \cite{DFFJM2009}-\cite{DY2008}, \cite{FMM2004}, \cite{GFM2006}-\cite{GMY2009}, \cite{YTUP2011}-\cite{ZTPP2011}, and the references therein.

Compared with the traditional periodic Hamming correlation, the partial Hamming correlation of FHSs is much less well studied. Nevertheless, in many application scenarios where the synchronization time is limited or the hardware is
complex \cite{EJHS2004}, the length of a correlation window should be much shorter than the period of the chosen FHSs  \cite{EJHS2004}. Therefore, the partial Hamming correlation, rather than the periodic Hamming correlation, will paly a major role in determining the performance.

In recent years, a little progress on the study of the partial Hamming correlation of FHSs has been made. In 2004, Eun et al. \cite{EJHS2004} generalized the Lempel-Greenberger bound on the periodic Hamming autocorrelation to the case of partial Hamming autocorrelation, and obtained a class of FHSs with optimal partial autocorrelation \cite{US1998}. In 2012, Zhou et al. \cite{ZTNP2012} extended the Peng-Fan bounds on the periodic Hamming correlation of FHS sets to the case of partial Hamming correlation. Based on the so-called array structure, Zhou et al. \cite{ZTNP2012} constructed both individual FHSs and FHS sets with optimal partial Hamming correlation. In 2014, Cai et al. \cite{CZYT2014} improved lower bounds on the partial Hamming correlation of FHSs and FHS sets, and based on generalized cyclotomy, they constructed FHS sets with optimal partial Hamming correlation. Very recently, Cai et al. \cite{CYZT2016} derived upper bounds on the family sizes of FHS sets with respect to partial Hamming correlation from some classical bounds on error-correcting codes, and they presented strictly optimal FHS sets having optimal family sizes with respect to one of the new bounds. Fan et al. \cite{FCT2016} established a generic connection between strictly optimal FHSs and disjoint cyclic perfect Mendelsohn difference families. Bao et al. \cite{BJFHS} established a correspondence between FHS sets with optimal partial Hamming correlation and multiple partition-type balanced nested cyclic difference packings with a special property. By virtue of this correspondence, they obtained some classes of strictly optimal FHSs and FHS sets by cyclotomic classes, they presented two recursive constructions for strictly optimal FHS sets and they also yielded some classes of strictly optimal FHS sets by these recursive constructions.

In this paper, we present some constructions for FHS sets with optimal partial Hamming correlation. First of all, We present several direct constructions for BNCDPs and BNCRDPs such that both of them have a special property by using trace functions and discrete logarithm. Next, we present three recursive constructions for FHS sets with partial Hamming correlation. Combing these BNCDPs, BNCRDPs and three recursive constructions, we yield infinitely many strictly optimal FHS sets with new and flexible parameters not covered in the literature. The parameters of FHS sets with optimal partial Hamming correlation from the known results and the new ones are listed in the Table I and Table II, respectively.

\newcounter{mytempeqncnt}
\begin{figure*}[!t]
\normalsize
\setcounter{mytempeqncnt}{\value{equation}}

\centerline{ TABLE I}
\vspace{0.2cm}

\centerline{\footnotesize SOME KNOWN FHS SETs WITH OPTIMAL PARTIAL HAMMING CORRELATION }
\vspace{0.2cm}
\begin{center}
\begin{tabular}{|c|c|c|c|c|c|}
\hline
Length & \tabincell{c} {Alphabet \\ size}& \tabincell{c}{$H_{max}$ over correlation\\ window of length L}& \tabincell{c} {Number of \\ sequences} & Constraints & Source \\
\hline
\tabincell{c}{$\frac{q^m-1}{d}$} & $q^{m-1}$ & $\left\lceil \frac{L(q-1)}{q^m-1}\right\rceil$ & $d$ & \tabincell{c}{ $d|(q-1)$,\\ gcd$(d,m)=1$ }& \cite{ZTNP2012} \\ \hline
\tabincell{c}{$ev$} & $v$ & $\left\lceil \frac{L}{v}\right\rceil$ & $f$ & \tabincell{c}{\\  \\} & \cite{CZYT2014} \\ \hline
\tabincell{c}{$ p(p^m-1)$ } & $p^{m}$ & $\left\lceil\frac{L}{p^{m}-1}\right\rceil$ & $p^{m-1}$ & \tabincell{c}{ \\ \tabincell{c}{$m\geq 2$} \\  } & \cite{CYZT2016} \\ \hline
\tabincell{c}{\\ $evw$ \\ \\ } & $(v-1)w+\frac{ew}{r}$& $\left\lceil \frac{L}{vw}\right\rceil$ & $f$ &  \tabincell{c}{ $q_1\geq p_1>2e$, \\$v\geq \frac{p_1e}{r}$ and $gcd(w,e)=1$ }  & \cite{BJFHS} \\ \hline
\tabincell{c}{$v\frac{q^m-1}{d}$ } & $vq^{m-1} $ & $\left\lceil \frac{(q-1)L}{v(q^{m}-1)}\right\rceil$ & $d$ &\tabincell{c}{ $m>1,$ $q^m \leq p_1$ and $gcd(d,m)=1,$\\$d|q-1,\frac{q^m-1}{d}|p_i-1$ for $1\leq i \leq s$  }&  \cite{BJFHS} \\ \hline
\end{tabular}
\end{center}

\hspace{1cm} $q$ is a prime power and $p$ is a prime;

\hspace{1cm} $v$ is an integer with prime factor decomposition $v=p_1^{m_1}p_2^{m_2}\cdots p_s^{m_s}$ with $p_1<p_2<\cdots <p_s$;

\hspace{1cm} $e,f$ are integers such that $e>1$ and $e|gcd(p_1-1,p_2-1,\ldots,p_s-1)$, and $f=\frac{p_1-1}{e}$;

\hspace{1cm} $w$ is an integer with prime factor decomposition $w=q_1^{n_1}q_2^{n_2}\cdots q_t^{n_t}$ with $q_1<q_2<\cdots <q_t$;

\hspace{1cm} $r$ is an integer such that $r>1$ and $r|gcd(e,q_1-1,q_2-1,\ldots,q_t-1)$;

\hspace{1cm} $d,m$ are positive integers.
\setcounter{equation}{\value{mytempeqncnt}}
\vspace*{2pt}
\end{figure*}

{ \footnotesize
\begin{figure*}[!t]
\small
\setcounter{mytempeqncnt}{\value{equation}}

\centerline{ TABLE II}
\vspace{0.2cm}

\centerline{\footnotesize NEW FHS SETS WITH OPTIMAL PARTIAL HAMMING CORRELATION }
\vspace{0.2cm}
\begin{center}
\begin{tabular}{|c|c|c|c|c|c|}
\hline
Length & \tabincell{c} {Number of \\ sequences} & $H_{max}$ & \tabincell{c} {Alphabet \\ size} & Constraints & Reference \\
\hline
\tabincell{c}{$\frac{w(q^m-1)}{d}$ } & $d$  & $\frac{q-1}{d}$ & $(q^{m-1}-1+\frac{q-1}{d})w$ & \tabincell{c}{  $m\geq 3$,\ $d|q-1$, \\ gcd$(m,d)=1$,\ $q_1>q,$}   & Corollary\ \ref{d3} \\ \hline
\tabincell{c}{$wp(p^m-1)$ } & $p^{m-1}$  & $p$ & $p^{m}w$ & \tabincell{c}{$m>1,\ $ \\ and $q_1>p^m,$ }   & Corollary\ \ref{uv2}\\ \hline
\tabincell{c}{$ewv$}  & $f$& $e$ & $(v-1+e)w$  &  \tabincell{c}{$q_1>p_1-1$,\\If $v$ is not a prime with $f>1$ \\ or $v$ is a prime with $f\geq e$,} &  Corollary\ \ref{uv5} \\ \hline
\tabincell{c}{$p\prod\limits_{i=1}^s(p^{u_i}-1)$ } & $p^{u_1-1}$  & $p$ & $1+\sum\limits_{i=1}^s p^{u_i}p^{u_{i+1}}\ldots p^{u_s}$ & \tabincell{c}{$p^{u_2}\geq p^{3u_1+1}u_2$, \\ $s\geq 3$,  $p^{u_2}\geq 2^s$, \\ $p^{u_1-1}\geq 5$}   & Corollary\ \ref{optimal p 4} \\ \hline
\tabincell{c}{$p(2^{p^m}-1)(\prod\limits_{i=1}^s(p^{u_i}-1))$ } & $p^{u_1-1}$  & $p$ & $2^{p^m}(1+\sum\limits_{i=1}^s\prod\limits_{j=i}^s p^{u_j})-1$ & \tabincell{c}{ $p^{u_2}\geq p^{3u_1+1}u_2$, \\ $s\geq 3$, $p^{u_2}\geq 2^s$, \\ $p^{u_1-1}\geq 5$,\\ $2^{p^m}>p^{3u_1}-p^{2u_1}$}   & Corollary\ \ref{optimal p 5} \\ \hline
\tabincell{c}{$pv(p^m-1)$}& $f_1$ & $p$ & $vp^m$ & \tabincell{c}{$p|p_i-1$, $f_1=\frac{p_1-1}{p}\geq 2$,\\ $p^m >p_1-1$, \\$p^m>2(1+p)$ }  & Corollary\ \ref{euv} \\ \hline
\tabincell{c}{$(q'-1)\frac{q^m-1}{d}$ } & $d$  & $\frac{q-1}{d}$ & $(q^{m-1}-1+\frac{q-1}{d})q'$ & \tabincell{c}{  $d|q-1$, gcd$(m,d)=1$, \\ $gcd(q'-1,\frac{q-1}{d})=1$,\\ $d\geq 2$, $q'>q+1$ }   & Corollary\ \ref{d4} \\ \hline
\tabincell{c}{$(q-1)p(p^m-1)$}& $p^{m-1}$ &  $p$ & $p^{m}q$ & \tabincell{c}{ $m>1$, $q\geq p^m$, \\ $gcd(q-1,p)=1$,\\ $p^{m}-3p\geq 1$ } & Corollary\ \ref{uv4}\\ \hline

\tabincell{c}{$ev(q-1)$}& $f$&  $e$ & $(v-1+e)q$ & \tabincell{c}{ $q> p_1-1$, $v\geq e^3f^2$,\\$gcd(q-1,e)=1$, \\ $q\geq 2e+5$, $f>1$} & Corollary\ \ref{uv3} \\ \hline
\end{tabular}
\end{center}

\hspace{1cm} $u_1, u_2, \ldots u_s $ are positive integers such that $u_1\leq u_2\leq \ldots \leq u_s$ ;

\hspace{1cm} $q,q'$ are prime powers;


\hspace{1cm} $v$ is an integer with prime factor decomposition $v=p_1^{m_1}p_2^{m_2}\cdots p_s^{m_s}$ with $p_1<p_2<\ldots<p_s$;

\hspace{1cm} $e$ is an integer such that $e|gcd(p_1-1,p_2-1,\ldots,p_s-1)$, and $f=\frac{p_1-1}{e}$;

\hspace{1cm} $w$ is any an integer with prime factor decomposition $w=q_1^{n_1}q_2^{n_2}\cdots q_t^{n_t}$ with $q_1<q_2<\ldots<q_t$;

\hspace{1cm} $d,m$ are positive integers.
\setcounter{equation}{\value{mytempeqncnt}}
\vspace*{2pt}
\end{figure*}
}

The outline of the paper is as follows. Section II introduces the known bounds on the partial Hamming correlation of FHSs and FHS sets. Section III presents several direct constructions for BNCDPs and BNCRDPs such that both of them have a special property by using trace functions and discrete logarithm. Section IV presents three recursive constructions of FHS sets with partial Hamming correlation. Section V concludes this paper with some remarks.

\section{Lower bounds on the partial Hamming correlation of FHSs and FHS sets} %
\label{pre}                     %

In this section, we introduce some known lower bounds on the partial Hamming correlation of FHSs and FHS sets.

For any positive integer $l\geq2$, let $F=\{f_0, f_1,\ldots, f_{l-1}\}$ be a set of $l$ available frequencies, also called an {\em alphabet}. A sequence $X=\{x(t)\}_{t=0}^{n-1}$ is called a {\em frequency hopping sequence} (FHS) of
length $n$ over $F$ if $x(t)\in F$ for all $0\leq t\leq n-1$. For any two FHSs  $X=\{x(t)\}_{t=0}^{n-1}$ and  $Y=\{y(t)\}_{t=0}^{n-1}$ of length $n$ over $F$, the {\em partial Hamming correlation} function of $X$ and $Y$ for a correlation window length $L$ starting at $j$ is defined by
\begin{equation}
\label{Correlation}
H_{X,Y}(\tau;j|L)=\sum_{t=j}^{j+L-1}h[x(t), y(t+\tau)],
\end{equation}
where $\tau,L,j$ are integers with $1\leq L\leq n$, $0\leq \tau, j< n$, $ h[a,b]=1$ if $a=b$ and $0$ otherwise, and the addition is performed modulo $n$. In particular, if $L=n$, the partial Hamming correlation function defined in ({\ref{Correlation}}) becomes the {\em conventional periodic  Hamming correlation} \cite{LG1974}.
If $x(t)=y(t)$ for all $0\leq t \leq n-1$, i.e., $X=Y$, we call $H_{X,X}(\tau;j|L)$ the {\em partial Hamming autocorrelation} of $X$; otherwise, we call $H_{X,Y}(\tau;j|L)$ the {\em partial Hamming cross-correlation} of $X$ and $Y$. For any two distinct sequences $X,Y$ over $F$ and any given integer $1\leq L\leq n$, we define
$$H(X;L)=\max\limits_{0\leq j < n}\max\limits_{1\leq \tau < n}\{H_{X,X}(\tau;j|L)\}$$
and
$$H(X,Y;L)=\max\limits_{0\leq j < n}\max\limits_{0\leq \tau < n}\{H_{X,Y}(\tau;j|L)\}.$$

For any FHS of length $n$ over an alphabet of size $l$ and each window length $L$ with $1\leq L\leq n$, Eun et al. \cite{EJHS2004} derived a lower bound, 
which is a generalization of the Lempel-Greenberger bound \cite{LG1974}. Recently, such a lower bound was improved by Cai et al. \cite{CZYT2014}.





Let $S$ be a set of $M$ FHSs of length $n$ over an alphabet $ F$ of size $l$. For any given correlation window length $L$, the maximum nontrivial partial Hamming correlation $H(S;L)$ of the sequence set $S$ is defined by
$$H(S;L)=\max \{\max\limits_{X\in{S}}H(X;L),\max\limits_{X,Y\in S,\ X\neq Y}H(X,Y;L)\}.$$

Throughout this paper, we use $(n,M,\lambda;l)$ to denote a set $S$ of $M$ FHSs of length $n$ over an alphabet $F$ of size $l$, where $\lambda=H(S;n)$, and we use $(n,\lambda;l)$ to denote an FHS $X$ of length $n$ over an alphabet $F$ of size $l$, where $\lambda=H(X;n)$.

When $L=n$, Peng and Fan \cite{PF2004} described the bounds on $H(S;n)$, which take into consideration the number of sequences in the set $S$.
In 2012, Zhou et al. \cite{ZTNP2012} extended the Peng-Fan bounds to the case of the partial Hamming correlation. They obtained $ H(S;L)\geq \left\lceil \frac{L}{n}\cdot \frac{(nM-l)n}{(nM-1)l}\right \rceil $ and $ H(S;L)\geq \left\lceil \frac{L}{n} \cdot \frac{2InM-(I+l)Il}{(nM-1)M} \right\rceil.$
Recently, such lower bounds were improved by Cai et al. \cite{CZYT2014}.

\begin{lemma}(\cite{CZYT2014}) \label{optimal set}
Let $S$ be a set of $M$ FHSs of length $n$ over $F$ of size $l$.
Define $I=\lfloor \frac{nM}{l}\rfloor$. Then, for an arbitrary window length $L$ with $1\leq L\leq n$,
\begin{equation}
\label{Bound 5}
H(S;L)\geq \left\lceil \frac{L}{n} \left\lceil\frac{(nM-l)n}{(nM-1)l}\right\rceil\right \rceil
\end{equation}
and
\begin{equation}
\label{Bound 6}
H(S;L)\geq \left\lceil \frac{L}{n} \left\lceil\frac{2InM-(I+1)Il}{(nM-1)M}\right\rceil \right\rceil.
\end{equation}
\end{lemma}

\begin{lemma}(\cite{CLLL2016}) \label{bounds34}
Let $S$ be a set of $M$ FHSs of length $n$ over $F$ of size $l$.
Define $I=\lfloor \frac{nM}{l}\rfloor$. If $nM\geq l$, then

$$ H(S;L)\geq  \left\lceil\frac{(nM-l)n}{(nM-1)l}\right\rceil
= \left\lceil\frac{2InM-(I+1)Il}{(nM-1)M}\right\rceil. $$
\end{lemma}

{\bf Remark:}  By the Lemma \ref{bounds34}, we have the two Peng-Fan bounds are identical.

Recall that the correlation window length may change from case to case according to the channel conditions in practical systems. Hence, it is very desirable that the involved FHS sets have optimal partial Hamming correlation for any window length. 
Cai et al. gave the following definition of strictly optimal FHS sets in \cite{CZYT2014}.

\begin{definition} \rm
\label{d111}
An FHS set $S$ is said to be {\em strictly optimal} or an {\em FHS set with optimal partial Hamming correlation} if the bounds in Lemma \ref{optimal set} is met for an arbitrary correlation window length $L$ with $1 \leq L \leq n.$
\end{definition}

When $L=n$, the bounds in Lemma \ref{optimal set} are exactly the Peng-Fan bounds. It turns out that each strictly optimal FHS set is also optimal with respect to the Peng-Fan bounds, but not vice versa.

\section{Strictly optimal FHS sets}

\subsection{A combinatorial characterization of Strictly optimal FHS sets} %

In 2009, Ge et al. \cite{GMY2009} revealed a connection between FHS sets and families of partition-type balanced nested cyclic difference packings.

Set $\mathbf{I}_l=\{0,1,\ldots,l-1\}$. Let $A$ be a subset of $\mathbb{Z}_v$, we define $m \cdot A=\bigcup\limits_{i=0}^{m-1}A$, where $\bigcup$ is the multiset union, i.e., the multiset $m \cdot A$ contains  $m$ copies of each element of $A$. Let $A, D$ be subsets of $\mathbb{Z}_v$, we define $ \Delta (D)\leq m \cdot A$, i.e., the multiset $\Delta (D)$ contains each element of $A$ at most $m$ times and no elements of $\mathbb{Z}_v \setminus A$ occur.  Let Tr$_{q^m/q}(x)$ be the trace function from $\mathbb{F}_{q^m}$ to $\mathbb{F}_q$.


Let $A,B$ be two subsets of $\mathbb{Z}_v$.  {\em The list of external difference} of ordered pair $(A,B)$ is the multiset $$\Delta_E(A,B)=\{y-x\colon (x,y)\in A\times B\}.$$  Note that the list of external difference $\Delta_E(A,B)$ may contain zero. For any residue $\tau\in \mathbb{Z}_v$, the number of occurrences of $\tau$ in $\Delta_E(A,B)$ is clearly equal to $|(A+\tau)\cap B|$.

Let ${\cal B}_j, 0\leq j\leq M-1$, be a collection of $l$ subsets $B_0^j, \ldots,B_{l-1}^j$ of $\mathbb{Z}_n$, respectively. The list of external difference of ordered pair $({\cal B}_j,{\cal B}_{j'})$, $0\leq j\neq j'<M$, is the union of multisets
$$\Delta_E({\cal B}_j,{\cal B}_{j'})= \bigcup\limits_{i\in \mathbf{I}_l} \Delta_E(B_i^j,B_i^{j'}).$$
If each ${\cal B}_j$ is an $(n,K_j,\lambda)$-CDP of size $l$, and $\Delta_E({\cal B}_j, {\cal B}_{j'})$
contains each residue of $\mathbb{Z}_n$ at most $\lambda$ times for $0\leq j\neq j'<M$, then the set $\{{\cal B}_0, \ldots, {\cal B}_{M-1}\}$ of CDPs is said to be {\em balanced nested} with index $\lambda$ and denoted by $(n,\{K_0,\ldots, K_{M-1}\},\lambda)$-BNCDP. If each ${\cal B}_j$ is a partition-type CDP for $0\leq j<M$, then the $(n,\{K_0,\ldots, K_{M-1}\},\lambda)$-BNCDP is called {\em partition-type}. For convenience, the number $l$ of the base blocks in ${\cal B}_j$ is also said to be the size of the BNCDP.

In 2009, Ge et al. \cite{GMY2009} revealed a connection between FHS sets and partition-type BNCDPs as follows.

\begin{theorem}{\rm(\cite{GMY2009})}\label{FHS set=DPs}
There exists an $(n,M,\lambda;l)$-FHS set over a frequency alphabet $F$ if and only if there exists a partition-type  $(n,\{K_0,K_1,\ldots,K_{M-1}\},\lambda)$-BNCDP of size $l$.
\end{theorem}

For a $u$-tuple $T=(a_0,a_1,\ldots,a_{u-1})$ over $\mathbb{Z}_n$, the multiset $\Delta_i(T)=\{a_{j+i}-a_j\colon 0\leq j\leq u-1\}$ is called $i$-apart difference list of the tuple, where $1\leq i\leq u$, $j+i$ is reduced modulo $u$, and $a_{j+i}-a_j$ is taken as the least positive residue modulo $n$.

Let ${\cal B}=\{{\cal B}_X\colon X\in S\}$ be a family of $M$ partition-type CDPs of size $l$ over $\mathbb{Z}_n$, where ${\cal B}_X=\{B_0^X,B_1^X,\ldots,B_{l-1}^X\}, \\ X \in S $. For two distinct partition-type CDPs ${\cal B}_X$, ${\cal B}_Y$, and for $0\leq \tau \leq n-1$, let
\[
\begin{array}{l}
\vspace{0.1cm}D_{(X,Y)}(\tau)=\{a\colon 0\leq a <n, \ (a,\ a+\tau) \in B_i^X\times B_i^Y, \ {\rm for\ some}\ i\in \mathbf{I}_l\},\\
\vspace{0.1cm}\overrightarrow{D_{(X,Y)}}(\tau)=(a_0,a_1,\ldots,a_{u-1}),\ {\rm where}\ 0\leq a_0<\cdots <a_{u-1}<n\ \  {\rm and}\ \{a_0,a_1,\ldots,a_{u-1}\}=D_{(X,Y)}(\tau), \\
 \end{array}
\]
\[
\begin{array}{l}
\vspace{0.1cm} d_i^{(X,Y)}=\min\limits_{0\leq \tau<n}\min \{z\colon z\in \{n\}\cup\Delta_i(\overrightarrow{D_{(X,Y)}}(\tau))\}, \ {\rm for}\ 0 \leq i\leq \max\{|D_{(X,Y)}(\tau)|\colon 0\leq \tau <n \},\\ {\rm and}\
\vspace{0.1cm} d_i^{{\cal B}}=\min\{ \min\limits_{X\neq Y}\{d_i^{(X,Y)} \}, \min\limits_{X \in S}\{d_i^{X}\}\},
 \end{array}
\]
where $\Delta_i(\overrightarrow{D_{(X,Y)}}(\tau))=\emptyset$ if $D_{(X,Y)}(\tau)=\emptyset$ or $i> |D_{(X,Y)}(\tau)|$. If $X=Y$, then $D_{(X,X)}(\tau), \overrightarrow{D_{(X,X)}}(\tau)$ and $d_i^{(X,X)}$ are the same as $D(\tau),\overrightarrow{D}(\tau)$ and $d^X_i$, respectively.
Note that $D_{(Y,X)}(-\tau)\equiv D_{(X,Y)}(\tau)+\tau \pmod n$ and $\min \{z\colon z\in \{n\}\cup \Delta_i(\overrightarrow{D_{(X,Y)}}(\tau))\}=\min \{z\colon z\in \{n\}\cup \Delta_i(\overrightarrow{D_{(Y,X)}}(-\tau))\}$.

In 2016, Bao et al. \cite{BJFHS} revealed a connection between strictly optimal FHS sets and partition-type BNCDPs with a special property as follows.
\begin{theorem}\label{optimal sets character}(\cite{BJFHS})
There is a strictly optimal $(n,M,\lambda;l)$-FHS set $S$ with respect to the Peng-Fan bounds if and only if there exists a partition-type $(n,\{K_X\colon X\in S\},\lambda)$-BNCDP of size $l$, ${\cal B}=\{{\cal B}_X=\{B_0^X,B_1^X,\ldots,B_{l-1}^X\}\colon X\in S\}$  over $\mathbb{Z}_n$ with $d_i^{{\cal B}} \geq\left\lfloor \frac{ni}{\lambda} \right \rfloor$ for $1\leq i\leq \lambda$, where
$K_X=\{|B_r^X|\colon 0\leq r\leq l-1\}$, $\lambda= \left\lceil\frac{2InM-(I+1)Il}{(nM-1)M}\right\rceil$ and $I= \left\lfloor\frac{nM}{l}\right\rfloor$.
\end{theorem}

Let ${\cal B}_j$ be an $(mg,g,K_j,1)$-CRDP over $\mathbb{Z}_{mg}$ for $0\leq j< M$, where ${\cal B}_j=\{B_0^j,B_1^j,\ldots,B_{u-1}^j\}$. The set $\{{\cal B}_0,\ldots,{\cal B}_{M-1}\}$ is referred to as an $(mg,g,\{K_0,K_1,\ldots,K_{M-1}\},1)$-BNCRDP ({\rm balanced nested cyclic relative difference packing}) over $\mathbb{Z}_{mg}$, if $\Delta({\cal B}_j,{\cal B}_{j'})$ contains each element of $\mathbb{Z}_{mg}\setminus m\mathbb{Z}_{mg}$ at most once and no elements of $m\mathbb{Z}_{g}$ occur for $0\leq j\neq j'<M$. For convenience, the number $u$ of the base blocks in ${\cal B}_j$ is also said to be the size of the BNCRDP.

One importance of a BNCRDP is that we can put an appropriate BNCRDP on its subgroup to derive a new BNCRDP.
\begin{lemma}\label{CRDPs=>CDPs 21}
Suppose that there exists an $(mg_1g_2,mg_1,\{K_0,\ldots,K_{M-1}\},1)$-BNCRDP of size $u_1$, $\{{\cal B}_0,\ldots,{\cal B}_{M-1}\}$, such that all elements of base blocks of ${\cal B}_j$, together with $0,g_2,\ldots,(\frac{mg_1}{s}-1)g_2$, form a complete system of representatives for the cosets of $\frac{mg_1g_2}{s}\mathbb{Z}_{mg_1g_2}$ in $\mathbb{Z}_{mg_1g_2}$ for $0\leq j < M$, where $s|m$. If there exists an $(mg_1, m, \{K_0',\ldots,K_{M-1}'\},1)$-BNCRDP of size $u_2$,
$\{{\cal B}'_0,\ldots,{\cal B}'_{M-1}\}$, such that all elements of base blocks of ${\cal B}'_j$, together with $0,g_1,\ldots,(\frac{m}{s}-1)g_1$, form a complete system of representatives for the cosets of $\frac{mg_1}{s}\mathbb{Z}_{mg_1}$ in $\mathbb{Z}_{mg_1}$ for $0\leq j < M$. Then there exists an
$(mg_1g_2, m,\{K_0\cup K_0',\ldots,K_{M-1}\cup K_{M-1}'\},1)$-BNCRDP of size $u_1+u_2$, $\{{\cal A}_0,\ldots,{\cal A}_{M-1}\}$, such that all elements of base blocks of ${\cal A}_j$, together with $0,g_1g_2,\ldots,(\frac{m}{s}-1)g_1g_2$, form a complete system of representatives for the cosets of $\frac{mg_1g_2}{s}\mathbb{Z}_{mg_1g_2}$ in $\mathbb{Z}_{mg_1g_2}$ for $0\leq j < M$.
\end{lemma}

\begin{IEEEproof}
Set ${\cal D}_i=\{g_2B:~  B\in {\cal B}'_i\}$ and ${\cal A}_i={\cal D}_i\cup {\cal B}_i$ for each $i$, $0\leq i<M$.

We show that $\{{\cal A}_0,\ldots,{\cal A}_{M-1}\}$ is an $(mg_1g_2, m,\{K_0\cup K_0',\ldots,K_{M-1}\cup K_{M-1}'\},1)$-BNCRDP of size $u_1+u_2$ such that all elements of base blocks of ${\cal A}_i$, together with $0,g_1g_2,\ldots,(\frac{m}{s}-1)g_1 g_2$, form a complete system of representatives for the cosets of $\frac{mg_1g_2}{s}\mathbb{Z}_{mg_1g_2}$ in $\mathbb{Z}_{mg_1g_2}$ for each $i$, $0\leq i<M$.

Since
\[
\begin{array}{l}
\vspace{0.2cm} \bigcup\limits_{B\in {\cal B}'_i}B \equiv \mathbb{Z}_{\frac{mg_1}{s}}\setminus  g_1\mathbb{Z}_{\frac{mg_1}{s}} \pmod {\frac{mg_1}{s}},
\end{array}
\]
it holds that
\[
\begin{array}{l}
\vspace{0.2cm} \bigcup\limits_{B\in {\cal D}_i}B =  \bigcup\limits_{B\in {\cal B}'_i}(g_2B) \  \equiv g_2\mathbb{Z}_{\frac{mg_1g_2}{s}}\setminus  g_1g_2\mathbb{Z}_{\frac{mg_1g_2}{s}} \pmod {\frac{mg_1g_2}{s}}.
\end{array}
\]
Since
\[
\begin{array}{l}
\vspace{0.2cm} \bigcup\limits_{B\in {\cal B}_i}B \equiv \mathbb{Z}_{\frac{mg_1g_2}{s}}\setminus  g_2\mathbb{Z}_{\frac{mg_1g_2}{s}} \pmod {\frac{mg_1g_2}{s}},
\end{array}
\]
we have that
\[
\begin{array}{l}
\vspace{0.2cm} \bigcup\limits_{B\in {\cal A}_i}B = (\bigcup\limits_{B\in {\cal B}_i}B)\cup (\bigcup\limits_{B\in {\cal D}_i}B) \  \equiv \mathbb{Z}_{\frac{mg_1g_2}{s}}\setminus g_1g_2\mathbb{Z}_{\frac{mg_1g_2}{s}} \pmod {\frac{mg_1g_2}{s}}.
\end{array}
\]
It follows that all elements of base blocks of ${\cal A}_i$, together with $0,g_1g_2,\ldots,(\frac{m}{s}-1)g_1 g_2$, form a complete system of representatives for the cosets of $\frac{mg_1g_2}{s}\mathbb{Z}_{mg_1g_2}$ in $\mathbb{Z}_{mg_1g_2}$ for each $i$, $0\leq i<M$.

For $0\leq i\not=j< M$, we show that $\Delta({\cal A}_i)$ and $\Delta_E({\cal A}_i, {\cal A}_j)$ contain each element of $ \mathbb{Z}_{mg_1g_2}\setminus g_1g_2 \mathbb{Z}_{mg_1g_2}$ at most once.

Since $\{{\cal B}'_0,\ldots,{\cal B}'_{M-1}\}$ is an $(mg_1, m, \{K_0',\ldots,K_{M-1}'\},$ $ 1)$-BNCRDP of size $u_2$ over $\mathbb{Z}_{mg_1}$ relative to $g_1\mathbb{Z}_{mg_1}$, we have that
\[
\begin{array}{l}
\vspace{0.2cm}\Delta({\cal B}'_i)\subset \mathbb{Z}_{mg_1}\setminus g_1\mathbb{Z}_{mg_1},\ {\rm and}\
\Delta_E({\cal B}'_i, {\cal B}'_j)\subset \mathbb{Z}_{mg_1}\setminus g_1\mathbb{Z}_{mg_1}.
\end{array}
\]
Then,
\[
\begin{array}{l}
\vspace{0.2cm}\Delta({\cal D}_i)\subset g_2\mathbb{Z}_{mg_1g_2}\setminus g_1g_2\mathbb{Z}_{mg_1g_2},\ {\rm and}\
\Delta_E({\cal D}_i, {\cal D}_j)\subset g_2\mathbb{Z}_{mg_1g_2}\setminus g_1g_2\mathbb{Z}_{mg_1g_2}.
\end{array}
\]
Since $\{{\cal B}_0,\ldots,{\cal B}_{M-1}\}$ is an $(mg_1g_2,mg_1,\{K_0,\ldots,K_{M-1}\},$ $1)$-BNCRDP of size $u_1$ over $\mathbb{Z}_{mg_1g_2}$ relative to $g_2\mathbb{Z}_{mg_1g_2}$, we have that
\[
\begin{array}{l}
\vspace{0.2cm}\Delta({\cal B}_i)\subset \mathbb{Z}_{mg_1g_2}\setminus g_2\mathbb{Z}_{mg_1g_2},\ {\rm and}\
\Delta_E({\cal B}_i, {\cal B}_j)\subset \mathbb{Z}_{mg_1g_2}\setminus g_2\mathbb{Z}_{mg_1g_2}.
\end{array}
\]
Hence,
\[
\begin{array}{l}
\vspace{0.2cm}\Delta({\cal A}_i)\subset \mathbb{Z}_{mg_1g_2}\setminus g_1g_2\mathbb{Z}_{mg_1g_2},\ {\rm and}\
\Delta_E({\cal A}_i, {\cal A}_j)\subset \mathbb{Z}_{mg_1g_2}\setminus g_1g_2\mathbb{Z}_{mg_1g_2}.
\end{array}
\]
Hence, $\Delta({\cal A}_i)$ and $\Delta_E({\cal A}_i, {\cal A}_j)$ contain each element of $ \mathbb{Z}_{mg_1g_2}\setminus g_1g_2\mathbb{Z}_{mg_1g_2}$ at most once.
This completes the proof.
\end{IEEEproof}

\begin{lemma}\label{CRDPs=>CDPs}(\cite{BJFHS})
Suppose that there exists an $(mg,g,\{K_0,\ldots,K_{M-1}\},1)$-BNCRDP of size $u$, $\{{\cal B}_0,\ldots,{\cal B}_{M-1}\}$, such that all elements of base blocks of ${\cal B}_j$, together with $0,m,\ldots,(s-1)m$, form a complete system of representatives for the cosets of $sm\mathbb{Z}_{mg}$ in $\mathbb{Z}_{mg}$ for $0\leq j < M$, where $s|g$. If there exists a partition-type $(g,\{K_0',\ldots,K_{M-1}'\},\frac{g}{s})$-BNCDP of size $l$, ${\cal A}$ such that $d_i^{\cal A}\geq si$ for $1\leq i\leq \frac{g}{s}$. Then there exists a partition-type $(mg,\{K_0\cup K_0',\ldots,K_{M-1}\cup K_{M-1}'\},\frac{g}{s})$-BNCDP of size $\frac{gu}{s}+l$, ${\cal D}$ such that $d_i^{\cal D} \geq ism$ for $1\leq i\leq \frac{g}{s}$.
\end{lemma}

When we replace the BNCRDP in Lemma \ref{CRDPs=>CDPs} with a $(g,\{K_0,K_1,\ldots,K_{M-1}\},1)$-BNCDP such that all elements of base blocks of each CDP form a complete system of representatives for the cosets of $s\mathbb{Z}_{g}$ in $\mathbb{Z}_{g}$ where $s|g$, the same procedure yields a new partition-type BNCDP. This proof is similar to that of Lemma 4.4 in \cite{BJFHS}.

\begin{lemma}\label{CRDPs=>CDPs 2}
Suppose that there exists a $(g,\{K_0,\ldots,K_{M-1}\},1)$-BNCDP of size $u$, ${\cal B}$, such that all elements of base blocks of each CDP form a complete system of representatives for the cosets of $s\mathbb{Z}_{g}$ in $\mathbb{Z}_{g}$ where $s|g$. Then there exists a partition-type $(g,\{K_0,\ldots,K_{M-1} \},\frac{g}{s})$-BNCDP of size $\frac{gu}{s}$, ${\cal D}$ such that $d_i^{\cal D} \geq is$ for $1\leq i\leq \frac{g}{s}$.
\end{lemma}

\begin{IEEEproof}
Let ${\cal B}=\{{\cal B}_i:~0\leq i<M\}$ and ${\cal D}_i=\{B+js:~ B\in {\cal B}_i, 0\leq j<{\frac{g}{s}}\}$ for each $i$, $0\leq i<M$. We show that ${\cal D}=\{{\cal D}_i:~0\leq i<M\}$ is a partition-type $(g,\{K_0,\ldots,K_{M-1} \},\frac{g}{s})$-BNCDP of size $\frac{gu}{s}$ such that $d_i^{\cal D} \geq is$ for $1\leq i\leq \frac{g}{s}$.

Since all elements of base blocks of ${\cal B}_i$ form a complete system of representatives for the cosets of $s\mathbb{Z}_{g}$ in $\mathbb{Z}_{g}$, we have that ${\cal D}_i$ is a partition of $\mathbb{Z}_{g}$.

Since ${\cal B}_i$ is a $(g,K_i,1)$-CDP over $\mathbb{Z}_{g}$ for each $i$, $0\leq i<M$, we have that $\Delta({\cal B}_i)\subset \mathbb{Z}_{g}\setminus \{0\}$ and $\Delta({\cal D}_i)\subset \frac{g}{s}\cdot (\mathbb{Z}_{g}\setminus \{0\}).$ Then, ${\cal D}_i$ is a $(g,K_i,\frac{g}{s})$-CDP over $\mathbb{Z}_{g}$ for each $i$, $0\leq i<M$.

Since $\{{\cal B}_i:~0\leq i<M\}$ is a $(g,\{K_0,\ldots,K_{M-1} \},1)$-BNCDP over $\mathbb{Z}_{g}$, we have that $\Delta_E({\cal B}_i, {\cal B}_j)\subset \mathbb{Z}_{g}$ and $\Delta_E({\cal D}_i, {\cal D}_j)\subset \frac{g}{s}\cdot \mathbb{Z}_{g}$ for $0\leq i\not= j<M.$ Hence, ${\cal D}$ is a partition-type $(g,\{K_0,\ldots,K_{M-1}\},\frac{g}{s})$-BNCDP of size $\frac{gu}{s}$.

From the construction, it is easy to see that if $\tau \in \Delta_E({\cal B}_i,{\cal B}_{j})$, then the orbit cycle $\overrightarrow{D_{(i,j)}}(\tau)$ of $\tau$ in $({\cal D}_i,{\cal D}_{j})$ is of the form $(a_0,a_0+s,a_0+2s,\ldots, a_0+g-s)$; otherwise ${D_{(i,j)}}(\tau)=\emptyset$. Then $d_a^{({\cal D}_i,{\cal D}_{j})}(\tau)=as$ for $\tau \in \Delta_E({\cal B}_i,{\cal B}_{j})$. Similarly, it is readily checked that $d_a^{{\cal D}_j}=as$ for $1\leq a\leq \frac{g}{s}$. Therefore, $d_a^{{\cal D}}=\min\{\min\limits_{0\leq i\neq j<M} \{d_a^{({\cal D}_i,{\cal D}_{j})}\}, \min\limits_{0\leq j<M} \{d_a^{{\cal D}_j}\}\}=as$ for $1\leq i \leq \frac{g}{s}$, and ${\cal D}$ is the required BNCDP. This completes the proof.
\end{IEEEproof}

\subsection{Some direct constructions of BNCDPs and BNCRDPs}
In this subsection, we give several direct constructions for BNCDPs and BNCRDPs such that both of them have a special property.

Let $d, m$ be positive integers such that $m\geq 2$ and gcd$(m,d)=1$, $q$ a prime power such that $d|q-1$. Let $\alpha$ be a primitive element of $\mathbb{F}_{q^m}$, set $g=\alpha^d$ and $\beta=\alpha^{\frac{q^m-1}{q-1}}$. Clearly, $\mathbb{F}_q^*=\{\beta^i:~ 0\leq i<q-1\}$.

Identifying $\mathbb{F}_{q^m}$ with the $m$-dimensional $\mathbb{F}_q$-vector space $\mathbb{F}_q^m$, then each element in $\mathbb{F}_{q^m}$ can be viewed as a vector over $\mathbb{F}_q$. Let $a_1,a_2,\ldots,a_{m-1} \in \mathbb{F}_{q^m}$ be $m-1$ linearly independent elements over $\mathbb{F}_q$. Define a sequence ${\bf v_i}=\{\overrightarrow{v_i(t)}\}_0^{\frac{q^m-1}{d}-1}$ of length $\frac{q^m-1}{d}$ over $\mathbb{F}_q^{m-1}$ as
\[
\begin{array}{l}
\overrightarrow{v_i(t)}=({\rm Tr}_{q^m/q}(\alpha^ia_1g^t),\ldots, {\rm Tr}_{q^m/q}(\alpha^ia_{m-1}g^t))
\end{array}
\]
where $0\leq t< \frac{q^m-1}{d}$ and $0\leq i<d$.

In 2012, Zhou et al. \cite{ZTNP2012} obtained the following result based on this construction.
\begin{theorem}\label{known result}(\cite{ZTNP2012})
Let $q$ be a prime power, $d, m$ positive integers such that $m\geq 2$, $d|q-1$ and gcd$(m,d)=1$. Then there is a strictly optimal $(\frac{q^m-1}{d},d,\frac{q-1}{d};q^{m-1})$-FHS set $S$, and $H(S;L)=\left\lceil \frac{L(q-1)}{q^m-1}\right\rceil$ for $1\leq L \leq \frac{q^m-1}{d}$.
\end{theorem}

For $\overrightarrow{b}=(b_1,b_2,\ldots, b_{m-1})\in \mathbb{F}_q^{m-1}$ and $0\leq i <d$, set
\[
\begin{array}{l}
\vspace{0.2cm}A_{\overrightarrow{b}}^i=\{t:~ {\rm Tr}_{q^m/q}(\alpha^ia_1g^t)=b_1, {\rm Tr}_{q^m/q}(\alpha^ia_2g^t)=b_2,\
 \ldots, {\rm Tr}_{q^m/q}(\alpha^ia_{m-1}g^t)=b_{m-1},\ 0\leq t<\frac{q^m-1}{d} \} \\
 {\rm and} \ {\cal A}_i=\{A_{\overrightarrow{b}}^i: \overrightarrow{b}\in \mathbb{F}_q^{m-1}\}.
\end{array}
\]

By Theorem \ref{optimal sets character} and Theorem \ref{known result}, $\{{\cal A}_i:0\leq i <d\}$ is a $(\frac{q^m-1}{d},\{K_0,\ldots,K_{d-1}\},\frac{q-1}{d} )$-BNCDP of size $q^{m-1}$.
Then, it holds that
\[
\begin{array}{l}
\vspace{0.2cm}\Delta ({\cal A}_i) \leq \frac{q-1}{d}\cdot (\mathbb{Z}_{\frac{q^m-1}{d}}\setminus\{0\}),\  {\rm and} \
\Delta ({\cal A}_i, {\cal A}_j) \leq \frac{q-1}{d} \cdot (\mathbb{Z}_{\frac{q^m-1}{d}})
\end{array}
\]
for $0\leq i\not= j<d$.

Since $\mathbb{F}_q^*=\{\beta^i:~ 0\leq i<q-1\}$, we have
{\footnotesize
\[
\begin{array}{l}
\vspace{0.2cm} j\frac{q^m-1}{q-1}+A_{\overrightarrow{b}}^i\\
\vspace{0.2cm} =\{j\frac{q^m-1}{q-1}+t:~ {\rm Tr}_{q^m/q}(\alpha^ia_1g^t)=b_1, {\rm Tr}_{q^m/q}(\alpha^ia_2g^t)=b_2,\
\ldots, {\rm Tr}_{q^m/q}(\alpha^ia_{m-1}g^t)=b_{m-1} ,\ 0\leq t<\frac{q^m-1}{d} \} \\
\vspace{0.2cm} =\{t:~ {\rm Tr}_{q^m/q}(\alpha^ia_1g^{t-j\frac{q^m-1}{q-1}})=b_1, {\rm Tr}_{q^m/q}(\alpha^ia_2g^{t-j\frac{q^m-1}{q-1}})=b_2,\  \ldots,{\rm Tr}_{q^m/q}(\alpha^ia_{m-1}g^{t-j\frac{q^m-1}{q-1}})=b_{m-1} ,\ 0\leq t<\frac{q^m-1}{d} \} \\
\vspace{0.2cm}=\{t:~ {\rm Tr}_{q^m/q}(\alpha^ia_1g^{t})=\beta^{jd}b_1, {\rm Tr}_{q^m/q}(\alpha^ia_2g^{t})=\beta^{jd}b_2,\  \ldots, {\rm Tr}_{q^m/q}(\alpha^ia_{m-1}g^{t})=\beta^{jd}b_{m-1} ,\ 0\leq t<\frac{q^m-1}{d} \} \\
\vspace{0.2cm} =A_{\beta^{jd}\overrightarrow{b}}^i
\end{array}
\]
}
where $\beta^{jd}\overrightarrow{b}=(\beta^{jd}b_1,\beta^{jd}b_2,\ldots,\beta^{jd} b_{m-1})\in \mathbb{F}_q^{m-1}$ for $\overrightarrow{b}=(b_1,b_2,\ldots, b_{m-1})\in \mathbb{F}_q^{m-1}$ and $0\leq j< \frac{q-1}{d}$.

From the proof of Theorem 2 in \cite{ZTNP2012}, we have that $A_{\overrightarrow{0}}^i\not= \emptyset$ for each $i$, $0\leq i<d$. Hence, there exist an element $s_i\in \mathbb{Z}_{\frac{q^m-1}{d}}$ such that $A_{\overrightarrow{0}}^i =\{s_i+j\frac{q^m-1}{q-1}:~ 0\leq j< \frac{q-1}{d}\}$ for each $i$, $0\leq i<d.$

Denote $G=\{1,\beta^d,\ldots, \beta^{q-1-d}\}$. Then $G$ is a multiplicative cyclic subgroup of order $\frac{q-1}{d}$ of $\mathbb{F}_q^*$. For $\overrightarrow{x},\overrightarrow{y}\in \mathbb{F}_q^{m-1}\setminus\{\overrightarrow{0}\}$, the binary relation $\sim$ defined by $\overrightarrow{x}\sim \overrightarrow{y}$ if and only if there exists a $g' \in G$ such that $g'\overrightarrow{x}=\overrightarrow{y}$ is an equivalence relation over $\mathbb{F}_q^{m-1}\setminus\{\overrightarrow{0}\}$. Then its equivalence classes are the subsets $G\overrightarrow{x}, \overrightarrow{x}\in \mathbb{F}_q^{m-1}\setminus\{\overrightarrow{0}\}$ of $\mathbb{F}_q^{m-1}$, where $G\overrightarrow{x}=\{g'\overrightarrow{x}:~ g'\in G\}$. Denote by $R$ a system of distinct representatives for the equivalence classes modulo $G$ of $\mathbb{F}_q^{m-1}\setminus\{\overrightarrow{0}\}$, then $|R|=\frac{d(q^{m-1}-1)}{q-1}$.

{\bf Construction A}\ \
For $0\leq i<d$ and $\overrightarrow{b}\in \mathbb{F}_q^{m-1}$, set
\[
\begin{array}{l}
\vspace{0.2cm}B_{\overrightarrow{b}}^i=\{a-s_i:~ a\in A_{\overrightarrow{b}}^i \}\ { \rm and} \\
{\cal A}'_i=\{B_{\overrightarrow{b}}^i: \overrightarrow{b} \in R \}
\end{array}
\]
where $A_{\overrightarrow{0}}^i =\{s_i, s_i+\frac{q^m-1}{q-1}, \ldots, s_i+\frac{(q^m-1)(q-1-d)}{(q-1)d} \}$ for each $i$, $0\leq i<d.$

Clearly,
{\footnotesize
{
\begin{equation}
\label{A}
\begin{aligned}
{\cal A}_i=(\bigcup\limits_{A \in {\cal A}'_i }\bigcup\limits_{0\leq j<\frac{q-1}{d}}\{A+s_i+j\frac{q^m-1}{q-1}\})
\cup\{A_{\overrightarrow{0}}^i\}.
\end{aligned}
\end{equation}
}
}
For $\overrightarrow{b}=(b_1,b_2,\ldots, b_{m-1})\in \mathbb{F}_q^{m-1}$, we have
{ \small
\[
\begin{array}{l}
\vspace{0.2cm}\sum \limits_{i=0}^{d-1} |B_{\overrightarrow{b}}^i|=\sum \limits_{i=0}^{d-1} |A_{\overrightarrow{b}}^i|\\
\vspace{0.2cm}\hspace{1.3cm}=\sum \limits_{i=0}^{d-1}|\{t:~ {\rm Tr}_{q^m/q}(\alpha^ia_1g^t)=b_1, {\rm Tr}_{q^m/q}(\alpha^ia_2g^t)=b_2,\ \ldots, {\rm Tr}_{q^m/q}(\alpha^ia_{m-1}g^t)=b_{m-1}, 0\leq t<\frac{q^m-1}{d} \}|\\
\vspace{0.2cm}\hspace{1.3cm}=|\{t:~ {\rm Tr}_{q^m/q}(a_1\alpha^t)=b_1, {\rm Tr}_{q^m/q}(a_2\alpha^t)=b_2,\  \ldots, {\rm Tr}_{q^m/q}(a_{m-1}\alpha^t)=b_{m-1}, 0\leq t<q^m-1 \}|.
\end{array}
\]
}
Then,
{
\begin{equation}
\label{sum}
\begin{aligned}
\sum \limits_{i=0}^{d-1} |B_{\overrightarrow{b}}^i|=
\left\{\begin{array}{ll}
q & {\rm  if} \ \ \overrightarrow{b}\not= \overrightarrow{0},\ \ {\rm and} \\
q-1  & {\rm otherwise}.
\end{array}
\right .
\end{aligned}
\end{equation}
}

\begin{theorem}\label{d}
Let\ $q$ be a prime power, $d, m$ positive integers such that $m\geq 2$, $d|q-1$ and gcd$(m,d)=1$. Let $\{{\cal A}'_i:~ 0\leq i<d\}$ be defined in {\bf Construction A}. Then $\{{\cal A}'_i:~ 0\leq i<d\}$ is a $(\frac{q^m-1}{d},\frac{q-1}{d}, \{K_0,\ldots, K_{d-1}\}, 1)$-BNCRDP of size  $\frac{d(q^{m-1}-1)}{q-1}$ such that all elements of base blocks of each CRDP, together with $0$, form a complete system of representatives for the cosets of $\frac{q^{m}-1}{q-1} \mathbb{Z}_{\frac{q^{m}-1}{d}}$ in $\mathbb{Z}_{\frac{q^{m}-1}{d}}$.
\end{theorem}

\begin{IEEEproof}
For $0\leq i< d$,
\[
\begin{array}{l}
\vspace{0.2cm}\bigcup\limits_{A \in {\cal A}_i } A=\mathbb{Z}_{\frac{q^{m}-1}{d}}.
\end{array}
\]
In view of equality (\ref{A}), it holds that
\[
\begin{array}{l}
\vspace{0.2cm}\bigcup\limits_{A \in {\cal A}_i } A=(\bigcup\limits_{A \in {\cal A}'_i }\bigcup\limits_{0\leq j<\frac{q-1}{d}}(A+s_i+j\frac{q^m-1}{q-1})) \cup A_0\\
\vspace{0.2cm}\hspace{1.2cm}=(\bigcup\limits_{0\leq j<\frac{q-1}{d}}(\bigcup\limits_{A \in {\cal A}'_i }(A+s_i+j\frac{q^m-1}{q-1}))) \cup A_0 \\
\vspace{0.2cm}\hspace{1.2cm}\equiv \frac{q-1}{d}\cdot ((\bigcup\limits_{A \in {\cal A}'_i }(A+s_i))\cup \{s_i\}) \pmod {\frac{q^{m}-1}{q-1}}\\
\vspace{0.2cm}\hspace{1.2cm}\equiv \frac{q-1}{d}\cdot ((\bigcup\limits_{A \in {\cal A}'_i }A)\cup \{0\}) \pmod {\frac{q^{m}-1}{q-1}}
\end{array}
\]
for $0\leq i< d$. Then,
\[
\begin{array}{l}
\bigcup\limits_{A \in {\cal A}'_i }A\equiv \{1,2,\ldots,\frac{q^{m}-1}{q-1}-1 \}\pmod {\frac{q^{m}-1}{q-1}}.
\end{array}
\]
It follows that all elements of base blocks of ${\cal A}'_i$, together with $0$, form a complete system of representatives for the cosets of $\frac{q^{m}-1}{q-1} \mathbb{Z}_{\frac{q^{m}-1}{d}}$ in $\mathbb{Z}_{\frac{q^{m}-1}{d}}$.

It remains to show that $\{{\cal A}'_i:0\leq i <d\}$ is  a $(\frac{q^m-1}{d},\frac{q-1}{d}, \{K_0,\ldots, K_{d-1}\}, 1)$-BNCRDP of size  $\frac{d(q^{m-1}-1)}{q-1}$.

In view of equality (\ref{A}), we get
\[
\begin{array}{l}
\vspace{0.2cm}\Delta ({\cal A}_i)= \frac{q-1}{d} \cdot ( \Delta ({\cal A}'_i)\cup \{\frac{q^m-1}{q-1},2\frac{q^m-1}{q-1}, \ldots, \frac{(q^m-1)(q-1-d)}{(q-1)d}  \}),\\
\vspace{0.2cm}\Delta_E ({\cal A}_i, {\cal A}_j)=\bigcup\limits_{\overrightarrow{b}\in \mathbb{F}_q^{m-1}} \Delta_E (A_{\overrightarrow{b}}^i,A_{\overrightarrow{b}}^j)\\
\vspace{0.2cm}\hspace{1.8cm}=\bigcup\limits_{k=0}^{\frac{q-1}{d}-1}\bigcup\limits_{\overrightarrow{b}\in R}\Delta_E (A_{\beta^{dk}\overrightarrow{b}}^i,A_{\beta^{dk}\overrightarrow{b}}^j) \cup \Delta_E(A_{\overrightarrow{0}}^i,A_{\overrightarrow{0}}^j)\\
\vspace{0.2cm}\hspace{1.8cm}=\frac{q-1}{d}\cdot (\bigcup\limits_{\overrightarrow{b}\in R}\Delta_E (s_i+B_{\overrightarrow{b}}^i, s_j+B_{\overrightarrow{b}}^j))\cup \Delta_E(A_{\overrightarrow{0}}^i,A_{\overrightarrow{0}}^j)\\
\vspace{0.2cm}\hspace{1.8cm}=\frac{q-1}{d} \cdot ( s_j-s_i+(\Delta_E ({\cal A}'_i,{\cal A}'_j)\cup \{0,\frac{q^m-1}{q-1},2\frac{q^m-1}{q-1}, \ldots, \frac{(q^m-1)(q-1-d)}{(q-1)d}  \})),
\end{array}
\]
where $G=\{1,\beta^d,\ldots, \beta^{q-1-d}\}$. By Theorem \ref{optimal sets character}, we have
\[
\begin{array}{l}
\vspace{0.2cm}\Delta ({\cal A}_i) \leq \frac{q-1}{d}\cdot (\mathbb{Z}_{\frac{q^m-1}{d}}\setminus\{0\}), {\rm and} \
\Delta_E ({\cal A}_i, {\cal A}_j) \leq \frac{q-1}{d} \cdot (\mathbb{Z}_{\frac{q^m-1}{d}})
\end{array}
\]
for $0\leq i\not= j<d$. Then,
\[
\begin{array}{l}
\vspace{0.2cm}\Delta ({\cal A}'_i) \leq \mathbb{Z}_{\frac{q^m-1}{d}}\setminus \frac{q^m-1}{q-1}\mathbb{Z}_{\frac{q^m-1}{d}},\   {\rm and} \
\Delta_E ({\cal A}'_i,{\cal A}'_j) \leq \mathbb{Z}_{\frac{q^m-1}{d}}\setminus \frac{q^m-1}{q-1}\mathbb{Z}_{\frac{q^m-1}{d}}.
\end{array}
\]
This completes the proof
\end{IEEEproof}

Starting with a $(\frac{q^m-1}{d},\frac{q-1}{d}, \{K_0,\ldots, K_{d-1}\}, 1)$-BNCRDP in Theorem \ref{d2}. By adding a block $\{0\}$ to each CRDP, we obtain the following $(\frac{q^m-1}{d}, \{K_0,\ldots, K_{d-1}\}, 1)$-BNCDP.

\begin{corollary}\label{d2}
Let $q$ be a prime power, $d, m$ positive integers such that $d|q-1$, $m\geq 2$ and gcd$(m,d)=1$. Then there exists a $(\frac{q^m-1}{d}, \{K'_0,\ldots, K'_{d-1}\}, 1)$-BNCDP of size $\frac{d(q^{m-1}-1)}{q-1}+1$ such that all elements of base blocks of each CDP form a complete system of representatives for the cosets of $\frac{q^{m}-1}{q-1} \mathbb{Z}_{\frac{q^{m}-1}{d}}$ in $\mathbb{Z}_{\frac{q^{m}-1}{d}}$.
\end{corollary}

{\bf Construction B}\ \ Let $q$ be a prime power and let $\alpha$ be a primitive element. Using the discrete logarithm in $\mathbb{F}_q^*$, define a function from $\mathbb{F}_q^*$ to $\mathbb{Z}_{q-1}$ as
$$\epsilon(x)=log_{\alpha}(x).$$

Let $p$ be a prime and let $m$ be an integer with $m>1$. Let $\alpha$ be a primitive element of $\mathbb{F}_{p^m}$ and denote $R=\{\sum_{i=1}^{m-1}a_i\alpha^i\colon a_i\in \mathbb{Z}_p, 1\leq i<m\}$. For $y\in R$ and $x\in R$, set
\[
\begin{array}{l}
 \vspace{0.2cm}A_y^x=\{(\epsilon(y-x-b),b):~  \ b \in \mathbb{Z}_{p}\setminus \{y-x\} \} \ {\rm and } \\
 {\cal A}^x=\{A_y^x:~ y\in R\}.
\end{array}
\]
Clearly, for $y, x\in R$,
\[
\begin{array}{l}
\vspace{0.2cm}A_y^x=\{(\epsilon(y-x-b),b):~  \ b \in \mathbb{Z}_{p}\setminus \{y-x\} \}\\
=\{(a,b):~ \alpha^a+b+x=y,(a,b)\in \mathbb{Z}_{p^m-1} \times \mathbb{Z}_{p}\}.
\end{array}
\]
\begin{lemma}\label{ p(p^m-1) 1}
Let $p$ be a prime and let $m$ be an integer with $m>1$. Let ${\cal A}^x$ be defined in Construction B. Then $ \Delta({\cal A}^x)  \subset (\mathbb{Z}_{p^m-1}\setminus\{0\})\times \mathbb{Z}_{p}$ and $\Delta({\cal A}^x, {\cal A}^{x'}) \subset  (\mathbb{Z}_{p^m-1}\setminus\{0\})\times \mathbb{Z}_{p}$ for $x\not= x'\in R$.
\end{lemma}

\begin{IEEEproof}
For $x, x' \in R$ and $(c,d)\in \mathbb{Z}_{p^m-1}\times \mathbb{Z}_{p}$, set
\[
\begin{array}{l}
\vspace{0.2cm} N_{x,x'}((c,d))=\sum\limits_{y\in R}|(A_y^x+(c,d))\cap A_y^{x'}|\\
\vspace{0.2cm}=|\{(a,b):~(a,b)\in A_y^x,\ (a+c,b+d) \in A_y^{x'},y\in R \}| \\
\vspace{0.2cm}=|\{(a,b):~ \alpha^a+b+x=y, \ \alpha^{a+c}+(b+d)+x'=y,\  y\in R \}|.
 \end{array}
\]
Then,
\[
\begin{array}{l}
\vspace{0.2cm} N_{x,x'}((c,d))=|\{(a,b):~ \alpha^a(1-\alpha^{c})=d+x'-x, \ {\rm and}  \
\alpha^{a}+b+x\in R \}|.
 \end{array}
\]

According to the values of $x, x'$ and $(c,d)$, we distinguish six cases.

Case 1: $x'= x$, $c=0$ and $d=0$. Since $|\{z+i:~ 0\leq i<p\}\cap R|=1$ for each $z \in \mathbb{F}_{p^m}$, it holds that
\[
\begin{array}{l}
\vspace{0.2cm} N_{x,x}((0,0))=|\{(a,b):~ \alpha^a+b+x\in R\}|\  =p^m-1.
 \end{array}
\]

Case 2: $x'= x$, $c \neq 0$ and $d=0$. In this case $\alpha^{c}-1\not=0$. Then
\[
\begin{array}{l}
\vspace{0.2cm} N_{x,x}((c,0))=|\{(a,b):~ \alpha^a(1-\alpha^{c})=0, \  \alpha^a+b+x\in R\}|\
 =0.
 \end{array}
\]

Case 3: $x'= x$, $c=0$ and $d \neq 0$. In this case $\alpha^{c}=1$. Then
\[
\begin{array}{l}
\vspace{0.2cm} N_{x,x}((0,d))=|\{(a,b):~ 0=d, \ \alpha^a+b+x \in R\}|\ =0.
 \end{array}
\]

Case 4: $x'= x$, $c\not=0$ and $d\not=0$. In this case $\alpha^{c}\not=1$, there exists a unique element $a_c\in \mathbb{Z}_{p^m-1}$ such that $ \alpha^{a_c}(1-\alpha^{c})=d$. Since $|\{z+i:~ 0\leq i<p\}\cap R|=1$ for each $z \in \mathbb{F}_{p^m}$, it holds that
\[
\begin{array}{l}
\vspace{0.2cm} N_{x,x}((c,d))=|\{(a,b):~ \alpha^a(1-\alpha^{c})=d, \  \alpha^a+b+x \in R\}|\\
\vspace{0.2cm}  \hspace{1.8cm}=|\{(a_c,b):~  \alpha^{a_c}+b+x \in R\}|\\
  \hspace{1.8cm} =1.
 \end{array}
\]

Case 5: $x'\not= x$ and $c=0$. In this case $\alpha^{c}=1$. Since $x, x'\in R$, we have $d\not=x-x'$. Then
\[
\begin{array}{l}
\vspace{0.2cm} N_{x,x'}((0,d))=|\{(a,b):~ 0=d+x'-x,\  \alpha^a+b+x\in R\}|\ =0.
 \end{array}
\]

Case 6: $x'\not= x$ and $c\not=0$. In this case $\alpha^{c}\not=1$. Since $x, x'\in R$, we have $d\not=x-x'$. Then  there exists a unique element $a_c\in \mathbb{Z}_{p^m-1}$ such that $ \alpha^{a_c}(1-\alpha^{c})=d+x'-x$. Since $|\{z+i:~ 0\leq i<p\}\cap R|=1$ for each $z \in \mathbb{F}_{p^m}$, it holds that
\[
\begin{array}{l}
\vspace{0.2cm} N_{x,x'}((c,d))=|\{(a,b):~ \alpha^a(1-\alpha^{c})=d+x'-x, \ \alpha^a+b+x \in R\}|\\
\vspace{0.2cm}  \hspace{1.95cm}=|\{(a_c,b):~ \alpha^{a_c}+b+x \in R\}|\\
  \hspace{1.95cm} =1.
 \end{array}
\]

In summary, the discussion in the six cases above shows that $ \Delta({\cal A}^x) \subset (\mathbb{Z}_{p^m-1}\setminus\{0\})\times \mathbb{Z}_{p}$ and $\Delta({\cal A}^x, {\cal A}^{x'}) \subset  (\mathbb{Z}_{p^m-1}\setminus\{0\})\times \mathbb{Z}_{p}$.
\end{IEEEproof}

\begin{theorem}\label{p(p^m-1) 2}
Let $p$ be a prime and let $m$ be an integer with $m>1$. Let ${\cal A}^x$ be defined in Construction A. Then $\{{\cal A}^x:~ x\in R\}$ is a $(p(p^m-1),\{K_0,\ldots,K_{p^{m-1}-1}\},1 )$-BNCDP of size $p^{m-1}$ such that all elements of base blocks of each CDP form a complete system of representatives for the cosets of $(p^m-1) \mathbb{Z}_{p(p^m-1)}$ in $\mathbb{Z}_{p(p^m-1)}$ where $K_0=\cdots=K_{p^{m-1}-1}=\{p,p-1\}$.
\end{theorem}

\begin{IEEEproof}
By Lemma \ref{ p(p^m-1) 1}, ${\cal A}^x$ is  a $(p(p^m-1),\{K_0,\ldots,K_{p^{m-1}-1}\},1 )$-BNCDP.
It is left to show that all elements of base blocks of each CDP form a complete system of representatives for the cosets of $(p^m-1) \mathbb{Z}_{p(p^m-1)}$ in $\mathbb{Z}_{p(p^m-1)}$.

Since $|\{z+i:~ 0\leq i<p\}\cap R|=1$ for each $z \in \mathbb{F}_{p^m}$, it holds that
\[
\begin{array}{l}
\vspace{0.2cm}{\cal A}^x=\bigcup\limits_{y\in R}A_{y}^x  \\
\vspace{0.2cm}=\bigcup\limits_{y\in R}\{(a,b):~ \alpha^a+b+x=y, \ (a,b)\in  \mathbb{Z}_{p^m-1}\times \mathbb{Z}_{p} \} \\
\vspace{0.2cm}=\{(a,b):~ \alpha^a+b\in R, \ (a,b)\in  \mathbb{Z}_{p^m-1}\times \mathbb{Z}_{p} \} \\
\equiv \mathbb{Z}_{p^m-1} \times \{0\}\pmod { \{0\}\times\mathbb{Z}_{p}}.
\end{array}
\]
This completes the proof.
\end{IEEEproof}

Applying Lemma \ref{CRDPs=>CDPs 2} and Theorem \ref{optimal sets character}, we obtain the following corollary. In 2016, Cai et al. \cite{CYZT2016} obtained a $(p(p^m-1),p^{m-1},p; p^m)$-FHS set. It is easy to check that the $(p(p^m-1),p^{m-1},p; p^m)$-FHS set are strictly optimal with respect to the Peng-Fan bounds.

\begin{corollary}\label{optimal p 1}(\cite{CYZT2016})
Let $p$ be a prime and let $m$ be an integer with $m>1$. Then there exists a strictly optimal $(p(p^m-1),p^{m-1},p; p^m)$-FHS set with respect to the Peng-Fan bounds over the alphabet $\mathbb{F}_{p^m}$.
\end{corollary}

{\bf Remark:} In 2016, Cai et al. have constructed such a strictly optimal $(p(p^m-1),p^{m-1},p;p^{m})$-FHS set \cite{CYZT2016}. Comparing with their proof, ours seems simpler.

Let ${\cal A}^x$ be block sets defined in construction B, set ${\cal B}^x=\{A_x^x\setminus\{(0,p-1)\}:~ A_x^x \in {\cal A}^x\}$ for each $x\in R$. Since $gcd(p,p^m-1)=1$, we have that $\mathbb{Z}_{p(p^m-1)}$ is isomorphic to $\mathbb{Z}_p\times \mathbb{Z}_{p^m-1}$. By using Lemma \ref{ p(p^m-1) 1} and  Theorem \ref{p(p^m-1) 2}, we obtain the following $(p(p^m-1),p, \{K_0,\ldots, K_{p^{m-1}-1}\}, 1)$-NBCRDP.


\begin{corollary}\label{optimal p 3}
Let $p$ be a prime and let $m$ be an integer with $m>1$. Then there exists a $(p(p^m-1),p, \{K_0,\ldots,\\ K_{p^{m-1}-1}\}, 1)$-NBCRDP of size $p^{m-1}$ such that all elements of base blocks of each CRDP, together with $0$, form a complete system of representatives for the cosets of $(p^m-1) \mathbb{Z}_{p(p^m-1)}$ in $\mathbb{Z}_{p(p^m-1)}$ where $K_0=\ldots= K_{p^{m-1}-1}=\{p,p-2\}$.
\end{corollary}

\begin{lemma}\label{cyclotomic construction}(\cite{BJFHS})
Let $v$ be a positive integer of the form $v=p_1^{m_1}p_2^{m_2}\cdots p_s^{m_s}$ for $s$ positive integers $m_1,m_2,\ldots,m_s$ and $s$ distinct primes $ p_1,p_2,\ldots,p_s$. Let $u$ be a positive integer such that $gcd(u,v)=1$. Let $e$ be a common factor of $u, p_1-1, p_2-1, \ldots, p_s-1$ and $e>1$, and let $f=\min \{\frac{p_i-1}{e}\colon 1\leq i\leq s\}$. Then there exists a $(uv,u,\{K_0,\ldots,K_{f-1}\},1)$-BNCRDP of size $\frac{v-1}{e}$ such that all elements of base blocks of each CRDP, together with $0$, form a complete system of representatives for the cosets of $v \mathbb{Z}_{uv}$ in $\mathbb{Z}_{uv}$ where $K_0=\cdots=K_{f-1}=\{e\}$.
\end{lemma}

{\bf Construction C}\ \  Starting with a $(uv,u,\{K_0,\ldots,K_{f-1}\},1)$-BNCRDP in Lemma \ref{cyclotomic construction} where $K_0=\cdots=K_{f-1}=\{e\}$. By adding a block $\{0\}$ to each CRDP, we obtain the following corollary.

\begin{corollary}\label{uv1}
Let $v$ be a positive integer of the form $v=p_1^{m_1}p_2^{m_2}\cdots p_s^{m_s}$ for $s$ positive integers $m_1,m_2,\ldots,m_s$ and $s$ distinct primes $ p_1,p_2,\ldots,p_s$. Let $u$ be a positive integer such that $gcd(u,v)=1$. Let $e$ be a common factor of $u, p_1-1, p_2-1, \ldots, p_s-1$ and $e>1$, and let $f=\min \{\frac{p_i-1}{e}\colon 1\leq i\leq s\}$. Then there exists a $(uv,\{K_0,\ldots,K_{f-1}\},1)$-BNCDP of size $\frac{v-1}{e}+1$ such that all elements of base blocks of each CDP form a complete system of representatives for the cosets of $v \mathbb{Z}_{uv}$ in $\mathbb{Z}_{uv}$ where $K_0=\cdots=K_{f-1}=\{e,1\}$.
\end{corollary}

\section{three recursive constructions of strictly optimal FHS sets}

\subsection{Construction based on difference matrices}

In this subsection, three recursive constructions are used to construct strictly optimal individual FHSs and FHS sets. The first recursive construction is based on the cyclic difference matrix (CDM).

A $(w, t, 1)$-CDM is a $t \times  w$ matrix $D=(d_{ij})$ ($0 \le i \le t-1$, $0 \le j \le w-1)$ with entries from $\mathbb{Z}_w$ such that, for any two distinct rows $R_{r}$ and $R_h$, the vector difference $R_h-R_{r}$ contains every residue of $\mathbb{Z}_w$ exactly once. It is easy to see that the property
of a difference matrix is preserved even if we add any element of $\mathbb{Z}_w$ to all entries in any row or column of the difference matrix.
Then, without loss of generality, we can assume that all entries in the first row are zero. Such a difference matrix is said to be {\em normalized}. The $(w,t-1,1)$-CDM obtained from a normalized $(w,t,1)$-CDM by deleting the first row is said to be {\em homogeneous}. The existence of a homogeneous $(w,t-1,1)$-CDM is equivalent to that of a $(w,t,1)$-CDM. Observe that difference matrices have been extensively studied. A large number of known $(w, t, 1)$-CDMs are well documented in \cite{CD2007}. In particular, the multiplication table of the prime field $\mathbb{Z}_p$ is a $(p, p, 1)$-CDM. By using the usual product construction of CDMs, we have the following existence result.

\begin{lemma}
\label{CDM}({\rm \cite{CD2007}})
Let $w$ and $t$ be integers with $w\geq t\geq 3$. If $w$ is odd and the least prime factor of $w$ is not less than $t$, then there exists a $(w,t,1)$-CDM.
\end{lemma}


When we replace the BNCRDP in Theorem 5.2 in \cite{BJFHS} with a $(g,\{K_0,K_1,\ldots,K_{M-1}\},1)$-BNCDP such that all elements of base blocks of each CDP form a complete system of representatives for the cosets of $s\mathbb{Z}_{g}$ in $\mathbb{Z}_{g}$ where $s|g$, the same procedure yields a new partition-type BNCDP. This proof is similar to that of Theorem 5.2 in \cite{BJFHS}.

\begin{theorem}\label{difference matrix 1}
Assume that $\{{\cal B}_0,\ldots,{\cal B}_{M-1}\}$ is a $(g,\{K_0,K_1,\ldots,K_{M-1}\},1)$-BNCDP of size $u$ such that all elements of base blocks of ${\cal B}_j$ form a complete system of representatives for the cosets of $s\mathbb{Z}_{g}$ in $\mathbb{Z}_{g}$for $0\leq j <M$, where $s|g$ and ${\cal B}_j=\{B_0^j,B_1^j,\ldots,B_{u-1}^j\}$. If there exists a homogeneous $(w, t, 1)$-CDM over $\mathbb{Z}_w$ with $t=\max \limits_{0\leq k <u}\{\sum \limits_{j=0}^{M-1} |B_k^j|\}$ and $gcd(w,\frac{g}{s})=1$, then there also exists a $(gw,\{K_0,K_1,\ldots,K_{M-1}\},1)$-BNCDP of size $uw$, ${\cal B}'=\{{\cal B}'_0,\ldots,{\cal B}'_{M-1}\}$ such that all elements of base blocks of ${\cal B}_{j}'$ form a complete system of representatives for the cosets of $sw\mathbb{Z}_{gw}$ in $\mathbb{Z}_{gw}$ for $0\leq j <M$.

\end{theorem}

\begin{IEEEproof}
Let $\Gamma=(\gamma_{i,j})$ be a homogeneous $(w,t,1)$-CDM over $\mathbb{Z}_w$.
For each collection of the following $M$ blocks:
\[
\begin{array}{l}
\vspace{0.2cm} B_i^0=\{a_{i,0,1},\ldots,a_{i,0,k_0}\},
\\ \vspace{0.2cm} B_i^1=\{a_{i,1,k_0+1},\ldots,a_{i,1,k_1}\},
\\ \vspace{0.2cm} \hspace{1.5cm} \vdots
\\ \vspace{0.2cm} B_i^{M-1}=\{a_{i,M-1,k_{M-2}+1},\ldots,a_{i,M-1,k_{M-1}}\},
\end{array}
\]
where $0\leq i <u$, we construct the following $uw$ new blocks:
\[
\begin{array}{l}
 \vspace{0.2cm}B_{(i,k)}^j=\{a_{i,j,k_{j-1}+1}+g\gamma_{k_{j-1}+1,k},\ldots,a_{i,j,k_{j}}+g\gamma_{k_{j},k}\},
\end{array}
\]
where $0\leq j < M, 0\leq k < w$ and $k_{-1}=0$.
Set
\[
\begin{array}{l}
\vspace{0.2cm} {\cal B}'_j=\{B_{(i,k)}^j\colon 0\leq i<u, 0\leq k<w\}, {\rm and}\\
 {\cal B}'=\{{\cal B}'_j\colon 0\leq j < M \},
\end{array}
\]
 then the size of ${\cal B}'_j$ is $uw$ for $0\leq j<M$. It is left to show  that ${\cal B}'$ is the required $(gw,\{K_0,K_1,\ldots,K_{M-1}\},1)$-BNCDP.

Firstly, we show that all elements of base blocks of ${\cal B}'_j$ form a complete system of representatives for the cosets of $sw\mathbb{Z}_{gw}$ in $\mathbb{Z}_{gw}$. Since $gcd(w,\frac{g}{s})=1$, we have that $\{c\cdot \frac{g}{s}\colon 0\leq c<w\}\equiv\{0,1,\ldots,w-1\}\pmod w$. It follows that  $\{s\cdot \frac{cg}{s}\colon 0\leq c<w\}\equiv \{cs\colon 0\leq c<w\} \pmod {sw}.$ Clearly, $\bigcup_{0\leq k<w}B_{(i,k)}^j=\bigcup_{z\in B_{i}^j}\{z+cg\colon 0\leq c<w\}$. Since all elements of base blocks of ${\cal B}_j$, form a complete system of representatives for the cosets of $s\mathbb{Z}_{g}$ in $\mathbb{Z}_{g}$, we have that $\bigcup_{0\leq i<u}B_{i}^j\equiv \{0,1,2,\ldots,s-1\} \pmod {s}$ and
\[
\begin{array}{l}
\vspace{0.2cm} \bigcup\limits_{0\leq i<u}\bigcup\limits_{0\leq k<w}B_{(i,k)}^j =\bigcup\limits_{0\leq i<u}\bigcup\limits_{z\in B_{i}^j}\{z+cg\colon 0\leq c<w\}\\
\vspace{0.2cm} \hspace{2.8cm} \equiv \bigcup\limits_{0\leq i<u}\bigcup\limits_{z\in B_{i}^j}\{z+cs\colon 0\leq c<w\} \\
\vspace{0.2cm}\hspace{2.8cm} \equiv \bigcup\limits_{z\in \mathbf{I}_{s}}\{z+cs\colon 0\leq c<w\}\\
\vspace{0.2cm}\hspace{2.8cm} \equiv \mathbf{I}_{sw}\pmod {sw},\\
\end{array}
\]
as desired.

Secondly, we show that each ${\cal B}'_j$ is a $(gw,K_j,1)$-CDP.
Since ${\cal B}_j$ is a $(g,K_j,1)$-CRDP, we have $\Delta({\cal B}_j)\subset \mathbb{Z}_{g}\setminus \{0\}$. Simple computation shows that
\[
\begin{array}{l}
\Delta({\cal B}'_j)=\bigcup\limits_{0\leq i<u, \atop 0\leq k<w}\Delta(B_{(i,k)}^j) \\
=\bigcup\limits_{0\leq i<u}\{a-b+cg\colon a\neq b \in B_i^j, \ 0\leq c < w\}\\
=\bigcup\limits_{\tau\in \Delta({\cal B}_j)}(g\mathbb{Z}_{gw}+\tau)\subset \mathbb{Z}_{gw}\setminus\{0\}.
\end{array}
\]
It follows that ${\cal B}'_j$ is a $(gw,K_j,1)$-CDP.

Finally, we show  $\Delta_E({\cal B}'_j, {\cal B}'_{j'})\subset \mathbb{Z}_{gw}$ for $0\leq j\neq j' <M$.
Since $\Delta_E({\cal B}_j, {\cal B}_{j'})\subset \mathbb{Z}_{g}$, we get
\[
\begin{array}{l}
\Delta_E({\cal B}'_j, {\cal B}'_{j'})=\bigcup\limits_{0\leq i<u}\bigcup\limits_{0\leq k<w}\Delta_E(B_{(i,k)}^j, B_{(i,k)}^{j'}) \\
=\bigcup\limits_{0\leq i<u}\{b-a+cg\colon (a,b) \in B_i^j\times B_i^{j'}, \ 0\leq c < w\}\\
=\bigcup\limits_{\tau\in \Delta_E({\cal B}_j,{\cal B}_{j'})}(g\mathbb{Z}_{gw}+\tau)\subset \mathbb{Z}_{gw}.
\end{array}
\]
Therefore, ${\cal B}'$ is the required BNCDP.
This completes the proof.
\end{IEEEproof}

Combining Theorem \ref{difference matrix 1}, Lemma \ref{CRDPs=>CDPs 2} and Corollary \ref{d2} together, we obtain the following corollary.

\begin{corollary}\label{d3}
Let $q$ be a prime power, $d, m$ positive integers such that $d|q-1$,  $m\geq 3$ and gcd$(m,d)=1$. Let $w$ be an odd integer whose the least prime factor is greater than $q$. Then there exists a strictly optimal $(\frac{w(q^m-1)}{d}, d, \frac{q-1}{d}; (q^{m-1}-1+\frac{q-1}{d})w)$-FHS set with respect to the Peng-Fan bounds.
\end{corollary}

\begin{IEEEproof}
By using Corollary \ref{d2}, there exists a $(\frac{q^m-1}{d}, \{K_0,\ldots, K_{d-1}\}, 1)$-BNCDP of size $\frac{d(q^{m-1}-1)}{q-1}+1$ such that all elements of base blocks of each CDP form a complete system of representatives for the cosets of $\frac{q^{m}-1}{q-1} \mathbb{Z}_{\frac{q^{m}-1}{d}}$ in $\mathbb{Z}_{\frac{q^{m}-1}{d}}$. Since $w$ is an odd integer whose the least prime factor is greater than $q$, there exists a homogeneous $(w, q, 1)$-CDM over $\mathbb{Z}_w$. Since the least prime factor of $w$ is greater than $q>\frac{q-1}{d}$, we have gcd$(w,\frac{q-1}{d})=1$. In view of equation (\ref{sum}), we have $\max \limits_{0\leq k <\frac{d(q^{m-1}-1)}{q-1}+1}\{\sum \limits_{j=0}^{d-1} |B_k^j|\}= q$. By Theorem \ref{difference matrix 1} with $g=\frac{q^{m}-1}{d}$ and $s=\frac{q^{m}-1}{q-1}$ yields a $(\frac{w(q^{m}-1)}{d},\{K_0,\ldots,K_{d-1}\},1 )$-BNCDP of size $(\frac{d(q^{m-1}-1)}{q-1}+1)w$ such that all elements of base blocks of each CDP form a complete system of representatives for the cosets of $\frac{w(q^{m}-1)}{q-1} \mathbb{Z}_{\frac{w(q^{m}-1)}{d}}$ in $\mathbb{Z}_{\frac{w(q^{m}-1)}{d}}$. By applying Lemma \ref{CRDPs=>CDPs 2}, we obtain a partition-type $(\frac{w(q^{m}-1)}{d},\{K_0,\ldots,K_{d-1}\},\frac{q-1}{d})$-BNCDP of size $(q^{m-1}-1+\frac{q-1}{d})w$, ${\cal D}$ such that $d_i^{\cal D} \geq i\frac{w(q^{m}-1)}{q-1}$ for $1\leq i\leq p$.

Finally, we show $\left\lceil\frac{2InM-(I+1)Il}{(nM-1)M}\right\rceil=\frac{q-1}{d}$.

Since $m\geq 3$, we have
\[
\begin{array}{l}
\vspace{0.2cm}I=\left\lfloor \frac{nM}{l} \right \rfloor=\left\lfloor \frac{\frac{w(q^{m}-1)}{d} \cdot d}{(q^{m-1}-1+\frac{q-1}{d})w} \right \rfloor  =q-1.
\end{array}
\]

Since $m\geq 3$, we have $1\leq 1+\frac{q(q-1)}{d}-q\leq (q-1)^2$ and $0< \frac{w(q-1)(1+\frac{q(q-1)}{d}-q)-(q-1)}{(w(q^{m}-1)-1)d}<1$. Then, it holds that
\[
\begin{array}{l}
\vspace{0.2cm}\left\lceil\frac{2InM-(I+1)Il}{(nM-1)M}\right\rceil =\left\lceil\frac{2(q-1)\frac{w(q^{m}-1)}{d} \cdot d-q(q-1)(q^{m-1}-1+\frac{q-1}{d})w}{(\frac{w(q^{m}-1)}{d} \cdot d-1)d}\right\rceil \\
\vspace{0.2cm}\hspace{2.5cm}= \left\lceil\frac{(q-1)w(q^{m}-2+q-\frac{q(q-1)}{d}))}{(w(q^{m}-1)-1)d}\right\rceil\\

\vspace{0.2cm}\hspace{2.5cm}= \left\lceil \frac{q-1}{d}-\frac{w(q-1)(1+\frac{q(q-1)}{d}-q)-(q-1)}{(w(q^{m}-1)-1)d}\right\rceil \\
\hspace{2.5cm} = \frac{q-1}{d}.
\end{array}
\]

By Theorem \ref{optimal sets character}, ${\cal D}$ is a strictly optimal $(\frac{w(q^m-1)}{d}, d, \frac{q-1}{d}; (q^{m-1}-1+\frac{q-1}{d})w)$-FHS set with respect to the Peng-Fan bounds. This completes the proof.
\end{IEEEproof}

Combining Theorem \ref{difference matrix 1}, Lemma \ref{CRDPs=>CDPs 2} and Theorem \ref{p(p^m-1) 2} together, we get the following corollary.

\begin{corollary}\label{uv2}
Let $p$ be a prime and let $m$ be an integer with $m>1$. Let $w$ be an odd integer whose the least prime factor is greater than $p^m$. Then there exists a strictly optimal $(wp(p^m-1),p^{m-1}, p; p^{m}w)$-FHS set with respect to the Peng-Fan bounds.
\end{corollary}

\begin{IEEEproof}
By using Theorem \ref{p(p^m-1) 2}, there exists a $(p(p^m-1),\{K_0,\ldots,K_{p^{m-1}-1}\},1 )$-BNCDP of size $p^{m-1}$ such that all elements of base blocks of each CDP form a complete system of representatives for the cosets of $(p^m-1) \mathbb{Z}_{p(p^m-1)}$ in $\mathbb{Z}_{p(p^m-1)}$ where $K_0=\cdots=K_{p^{m-1}-1}=\{p,p-1\}$. Since $w$ is an odd integer whose the least prime factor is greater than $p^m$, there exists a homogeneous $(w, p^m, 1)$-CDM over $\mathbb{Z}_w$. Since the least prime factor of $w$ is greater than $p^m$, we have gcd$(w,p)=1$. Since $\max \limits_{0\leq k <p^{m-1}}\{\sum \limits_{j=0}^{p^{m-1}-1} |B_k^j|\}\leq p\cdot p^{m-1}= p^m$, by Theorem \ref{difference matrix 1} with $g=p(p^m-1)$ and $s=p^m-1$ yields a $(p(p^m-1)w,\{K_0,\ldots,K_{p^{m-1}-1}\},1 )$-BNCDP of size $p^{m-1}w$ such that all elements of base blocks of each CDP form a complete system of representatives for the cosets of $(p^m-1)w \mathbb{Z}_{p(p^m-1)w}$ in $\mathbb{Z}_{p(p^m-1)w}$ where $K_0=\cdots=K_{p^{m-1}-1}=\{p,p-1\}$. By applying Lemma \ref{CRDPs=>CDPs 2}, we obtain a partition-type $(p(p^m-1)w,\{K_0,\ldots,K_{p^{m-1}-1}\},p)$-BNCDP of size $p^mw$, ${\cal D}$ such that $d_i^{\cal D} \geq i(p^m-1)w$ for $1\leq i\leq p$.

Finally, we show $\left\lceil\frac{2InM-(I+1)Il}{(nM-1)M}\right\rceil=p$.

By definition, we have
\[
\begin{array}{l}
\vspace{0.2cm}I=\left\lfloor \frac{nM}{l} \right \rfloor=\left\lfloor \frac{wp(p^m-1) \cdot p^{m-1}}{p^{m}w} \right \rfloor =p^m-1.
\end{array}
\]
Since $m>1$, we have
\[
\begin{array}{l}
\vspace{0.2cm} w(p^m-1) p^{m}-1-(2pw(p^m-1)-p))\\
\vspace{0.2cm} =w(p^m-1)(p^m-2p)+p-1\\
\vspace{0.2cm}>0, \ {\rm and} \\
0<\frac{p(2w(p^m-1)-1)}{w(p^m-1) p^{m}-1}<1.
\end{array}
\]
Then, it holds that
\[
\begin{array}{l}
\vspace{0.2cm}\left\lceil\frac{2InM-(I+1)Il}{(nM-1)M}\right\rceil =\left\lceil\frac{2(p^m-1)wp(p^m-1) \cdot p^{m-1}-p^m(p^m-1)p^{m}w}{(wp(p^m-1) \cdot p^{m-1}-1)p^{m-1}}\right\rceil \\
\vspace{0.2cm}\hspace{2.5cm}= \left\lceil p-\frac{p(2w(p^m-1)-1)}{w(p^m-1) p^{m}-1}\right\rceil \\
\hspace{2.5cm} =p.
\end{array}
\]

By Theorem \ref{optimal sets character}, ${\cal D}$ is a strictly optimal $(wp(p^m-1),p^{m-1}, p; p^{m}w)$-FHS set with respect to the Peng-Fan bounds. This completes the proof.
\end{IEEEproof}

Combining Theorem \ref{difference matrix 1}, Lemma \ref{CRDPs=>CDPs 2} and Corollary \ref{uv1} together, we obtain the following corollary.

\begin{corollary}\label{uv5}
Let $v$ be a positive integer of the form $v=p_1^{m_1}p_2^{m_2}\cdots p_s^{m_s}$ for $s$ positive integers $m_1,m_2,\ldots,m_s$ and $s$ distinct primes $ p_1,p_2,\ldots,p_s$ with $p_1<p_2<\cdots<p_s$. Let $e$ be a common factor of $p_1-1, p_2-1, \ldots, p_s-1$ and $e>1$, and let $f=\min \{\frac{p_i-1}{e}\colon 1\leq i\leq s\}$. Let $w$ be an odd integer whose the least prime factor is greater than $p_1-1$. If $v$ is not a prime with $f>1$ or $v$ is a prime with $f\geq e$, then there exists a strictly optimal $(ewv,f, e; (v-1+e)w)$-FHS set with respect to the Peng-Fan bounds.
\end{corollary}

\begin{IEEEproof}
By Corollary \ref{uv1} with $u=e$, there exists an $(ev,\{K_0,\ldots,K_{f-1}\},1)$-BNCDP of size $\frac{v-1}{e}+1$ such that all elements of base blocks of each CDP form a complete system of representatives for the cosets of $v \mathbb{Z}_{ev}$ in $\mathbb{Z}_{ev}$ where $K_0=\cdots=K_{f-1}=\{e,1\}$. Since $w$ is an odd integer whose the least prime factor is greater than $p_1-1$, there exists a homogeneous $(w, p_1-1, 1)$-CDM over $\mathbb{Z}_w$. Since the least prime factor of $w$ is greater than $p_1-1 \geq e$, we have gcd$(w,e)=1$. Since $\max \limits_{0\leq k <\frac{v-1}{e}+1}\{\sum \limits_{j=0}^{f-1} |B_k^j|\}= p_1-1$, by Theorem \ref{difference matrix 1} with $g=ev$ and $s=v$ yields an $(evw,\{K_0,\ldots,K_{f-1}\},1 )$-BNCDP of size $(\frac{v-1}{e}+1)w$ such that all elements of base blocks of each CDP form a complete system of representatives for the cosets of $vw \mathbb{Z}_{evw}$ in $\mathbb{Z}_{evw}$ where $K_0=\cdots=K_{f-1}=\{e,1\}$. By applying Lemma \ref{CRDPs=>CDPs 2}, we obtain a partition-type $(evw,\{K_0,\ldots,K_{f-1}\},e)$-BNCDP of size $(v-1+e)w$, ${\cal D}$ such that $d_i^{\cal D} \geq ivw$ for $1\leq i\leq e$.

Finally, we show $\left\lceil\frac{2InM-(I+1)Il}{(nM-1)M}\right\rceil=e$.

If $v$ is not a prime, we have $v\geq p_1^2>e^2f^2$ and
\[
\begin{array}{l}
\vspace{0.2cm}I=\left\lfloor \frac{nM}{l} \right \rfloor=\left\lfloor \frac{evw \cdot f}{(v-1+e)w} \right \rfloor =\left\lfloor ef- \frac{ef(e-1)}{v-1+e} \right \rfloor =ef-1.
\end{array}
\]
Since $f>1$ and $v\geq p_1^2$, we have
\[
\begin{array}{l}
\vspace{0.2cm}v(f-1)>v\geq p_1^2>e^2f+1-ef-e,\ {\rm and}\\
0<\frac{ew(v+e^2f+1-ef-e)-e}{evw f-1}<1.
\end{array}
\]
Then, it holds that
\[
\begin{array}{l}
\vspace{0.2cm}\left\lceil\frac{2InM-(I+1)Il}{(nM-1)M}\right\rceil =\left\lceil\frac{2(ef-1)evw \cdot f-ef(ef-1)(v-1+e)w}{(evw \cdot f-1)f}\right\rceil \\
\vspace{0.2cm}\hspace{2.5cm}=\left\lceil\frac{ew(efv+ef+e-e^2f-v-1)}{evw \cdot f-1}\right\rceil \\
\vspace{0.2cm}\hspace{2.5cm}= \left\lceil e-\frac{ew(v+e^2f+1-ef-e)-e}{evw f-1}\right\rceil \\
\hspace{2.5cm} =e.
\end{array}
\]

If $v$ is a prime with $f\geq e$, we have $v=ef+1$ and
\[
\begin{array}{l}
\vspace{0.2cm}I=\left\lfloor \frac{nM}{l} \right \rfloor=\left\lfloor \frac{evw \cdot f}{(v-1+e)w} \right \rfloor =\left\lfloor ef-e+1+ \frac{e-1}{f+1} \right \rfloor =ef-e+1.
\end{array}
\]

Since $e>1$, $f\geq e$ and $w\geq p_1-1=ef$, we have
\[
\begin{array}{l}
\vspace{0.2cm} ew(e^2(f^2-f)+2ef+(e^2-3e+2))>ef, \\
\vspace{0.2cm}ef^3+f^2>e^2f^2+e^2-(e-2)ef-(3e-2),\ {\rm and}\\
0<\frac{ew(e^2f^2+e^2-(e-2)ef-(3e-2))-ef}{ew(ef^3+f^2)-f}<1.
\end{array}
\]
Since $v=ef+1$ , we have
\[
\begin{array}{l}
\vspace{0.2cm}\left\lceil\frac{2InM-(I+1)Il}{(nM-1)M}\right\rceil =\left\lceil\frac{2(ef-e+1)evw \cdot f-(ef-e+2)(ef-e+1)(v-1+e)w}{(evw \cdot f-1)f}\right\rceil \\
\vspace{0.2cm}\hspace{2.5cm}=\left\lceil\frac{(ef-e+1)(ef^2+e-2)ew}{(evw \cdot f-1)f}\right\rceil \\
\vspace{0.2cm}\hspace{2.5cm}= \left\lceil e-\frac{ew(e^2f^2+e^2-(e-2)ef-(3e-2))-ef}{ew(ef^3+f^2)-f}\right\rceil \\
\hspace{2.5cm}=e.
\end{array}
\]

By Theorem \ref{optimal sets character}, ${\cal D}$ is a strictly optimal $(ewv,f, e; (v-1+e)w)$-FHS set with respect to the Peng-Fan bounds. This completes the proof.
\end{IEEEproof}

\subsection{Constructions based on discrete logarithm}

Now, we present two recursive constructions for strictly optimal frequency hopping seuqences based on discrete logarithm.


$\epsilon(x)$ is defined in Section III-B.

\begin{lemma}\label{q-1 power}\cite{}
Let $q$ be a prime power. Let $a,b \in \mathbb{F}_q$ be two distinct elements, then
$$\{\epsilon(x-a)-\epsilon(x-b):~x\in \mathbb{F}_q\setminus\{a,b\} \}=\mathbb{Z}_{q-1}\setminus\{0\}.$$
\end{lemma}


\begin{theorem}\label{q-1 1}
Assume that $\{{\cal B}_0,\ldots,{\cal B}_{M-1}\}$ is an $(mg,g,\{K_0,K_1,\ldots,K_{M-1}\},1)$-BNCRDP of size $u$ such that all elements of base blocks of ${\cal B}_j$, together with $0,m,\ldots,(s-1)m,$ form a complete system of representatives for the cosets of $sm\mathbb{Z}_{mg}$ in $\mathbb{Z}_{mg}$  for $0\leq j <M$, where $s|g$ and ${\cal B}_j=\{B_0^j,B_1^j,\ldots,B_{u-1}^j\}$. Let $q$ be a prime power such that $q\geq\max \limits_{0\leq k <u}\{\sum \limits_{j=0}^{M-1} |B_k^j|\}$ and $gcd(q-1,\frac{g}{s})=1$, then there also exists an $(mg(q-1),g(q-1),\{K'_0,K'_1,\ldots,K'_{M-1}\},1)$-BNCRDP of size $uq$, ${\cal B}'=\{{\cal B}'_0,\ldots,{\cal B}'_{M-1}\}$ such that all elements of base blocks of ${\cal B}_{j}'$, together with $0,m,\ldots,(s(q-1)-1)m$, form a complete system of representatives for the cosets of $sm(q-1)\mathbb{Z}_{mg(q-1)}$ in $\mathbb{Z}_{mg(q-1)}$ for $0\leq j <M$ where $K'_i\subset K_i\cup \{k-1:~ k\in K_i\}$ for each $i$.
\end{theorem}

\begin{IEEEproof}
For $0\leq i<u$ and $0\leq j < M$, let $B_i^j=\{a_{i,j,1},\ldots,a_{i,j,|B_i^j|}\}$.
Since $q\geq \max \limits_{0\leq i <u}\{\sum \limits_{j=0}^{M-1} |B_i^j|\}$,  for each $i$ we take $\sum \limits_{j=0}^{M-1} |B_i^j|$ distinct elements $x_{i,j,c} \in \mathbb{F}_q$ ($0\leq j<M$ and $1\leq c\leq |B_i^j|$) and set
\[
\begin{array}{l}
 \vspace{0.2cm}\eta_i(a_{i,j,c})=x_{i,j,c}.
\end{array}
\]
Clearly, $\eta_i(a_{i,b,c})=\eta_i(a_{i,b',c'})$ if and only if $(b,c)=(b',c')$ for each $i$.

For $0\leq i<u$ and $0\leq j<M$, we construct the following $q$ new blocks:

\[
\begin{array}{l}
 \vspace{0.2cm}B_{(i,y)}^j=\{ x+mg \epsilon(y-\eta_i(x)):~ x\in B_i^j, \eta_i(x) \not= y \},
\\ \hspace{1.5cm}\vspace{0.2cm} {\rm where}\ y\in \mathbb{F}_q.
\end{array}
\]

Set
\[
\begin{array}{l}
\vspace{0.2cm} {\cal B}'_j=\{B_{(i,y)}^j\colon 0\leq i<u, y\in \mathbb{F}_q\}, {\rm and}\\
 {\cal B}'=\{{\cal B}'_j\colon 0\leq j < M \},
\end{array}
\]
It is left to show  that ${\cal B}'$ is the required $(mg(q-1),g(q-1),\{K'_0,K'_1,\ldots,K'_{M-1}\},1)$-BNCRDP.

Firstly, we show that all elements of base blocks ${\cal B}'_j$, together with $0,m,\ldots,(s(q-1)-1)m$, form a complete system of representatives for the cosets of $sm(q-1)\mathbb{Z}_{mg(q-1)}$ in $\mathbb{Z}_{mg(q-1)}$. Since $gcd(q-1,\frac{g}{s})=1$, we have that $\{c\cdot \frac{g}{s}\colon 0\leq c<q-1\}\equiv\{0,1,\ldots,q-2\}\pmod {q-1}$. It follows that  $\{ms\cdot \frac{cg}{s}\colon 0\leq c<q-1\}\equiv \{csm\colon 0\leq c<q-1\} \pmod {sm(q-1)}.$ Clearly,

 $$\bigcup_{y\in \mathbb{F}_q}B_{(i,y)}^j=\bigcup_{z\in B_{i}^j}\{z+cmg\colon 0\leq c<q-1\}.$$
Since all elements of base blocks of ${\cal B}_j$, together with $0,m,2m,\cdots,(s-1)m$, form a complete system of representatives for the cosets of $sm\mathbb{Z}_{mg}$ in $\mathbb{Z}_{mg}$, we have that $\bigcup_{0\leq i<u}B_{i}^j\equiv \{0,1,2,\ldots,sm-1\}\setminus \{0,m,\ldots,sm-m\} \pmod {sm}$ and
\[
\begin{array}{l}
\vspace{0.2cm} \bigcup\limits_{0\leq i<u}\bigcup\limits_{y\in \mathbb{F}_q}B_{(i,y)}^j \\
\vspace{0.2cm} =\bigcup\limits_{0\leq i<u}\bigcup\limits_{z\in B_{i}^j}\{z+cmg\colon 0\leq c<q-1\}\\
\vspace{0.2cm} \equiv \bigcup\limits_{0\leq i<u}\bigcup\limits_{z\in B_{i}^j}\{z+csm\colon 0\leq c<q-1\} \\
\vspace{0.2cm} \equiv \bigcup\limits_{z\in \mathbf{I}_{sm}\setminus \{0,m,\ldots,sm-m\}}\{z+csm\colon 0\leq c<q-1\}\\
\vspace{0.2cm} \equiv \mathbf{I}_{sm(q-1)}\setminus \{0,m,\ldots,sm(q-1)-m\}\pmod {sm(q-1)},\\
\end{array}
\]
as desired.

Secondly, we show that each ${\cal B}'_j$ is an $(mg(q-1),g(q-1),K_j,1)$-CRDP.
Since ${\cal B}_j$ is an $(mg,g,K_j,1)$-CRDP, we have $\Delta({\cal B}_j)\subset \mathbb{Z}_{mg}\setminus m\mathbb{Z}_{mg}$. By Lemma \ref{q-1 power}, it holds that
\[
\begin{array}{l}
\Delta({\cal B}'_j)=\bigcup\limits_{0\leq i<u, \atop y\in \mathbb{F}_q}\Delta(B_{(i,y)}^j) \\
=\bigcup\limits_{0\leq i<u}\{a-b+cmg\colon a\neq b \in B_i^j, \ 0< c < q-1\}\\
=\bigcup\limits_{\tau\in \Delta({\cal B}_j)}(mg\mathbb{Z}_{mg(q-1)}\setminus \{0\}+\tau)\subset \mathbb{Z}_{mg(q-1)}\setminus m\mathbb{Z}_{mg(q-1)}.
\end{array}
\]
It follows that ${\cal B}'_j$ is an $(mg(q-1),g(q-1),K_j,1)$-CRDP.

Finally, we show  $\Delta_E({\cal B}'_j, {\cal B}'_{j'})\subseteq \mathbb{Z}_{mg(q-1)}\setminus m\mathbb{Z}_{mg(q-1)}$ for $0\leq j\neq j' <M$.
Since $\Delta_E({\cal B}_j, {\cal B}_{j'})\subset \mathbb{Z}_{mg}\setminus m\mathbb{Z}_{mg}$, we get
\[
\begin{array}{l}
\Delta_E({\cal B}'_j, {\cal B}'_{j'})=\bigcup\limits_{0\leq i<u}\bigcup\limits_{y\in \mathbb{F}_q}\Delta(B_{(i,y)}^j, B_{(i,y)}^{j'}) \\
=\bigcup\limits_{0\leq i<u}\{b-a+cmg\colon (a,b) \in B_i^j\times B_i^{j'}, \ 0< c < q-1\}\\
=\bigcup\limits_{\tau\in \Delta_E({\cal B}_j,{\cal B}_{j'})}(mg\mathbb{Z}_{mg(q-1)}\setminus \{0\} +\tau)\subset \mathbb{Z}_{mg(q-1)}\setminus m\mathbb{Z}_{mg(q-1)}.
\end{array}
\]
Therefore, ${\cal B}'$ is the required BNCRDP.
This completes the proof.
\end{IEEEproof}

Combining Theorem \ref{q-1 1}, Lemma \ref{CRDPs=>CDPs 21} and Corollary \ref{optimal p 3} together, we obtain the following corollary.

\begin{corollary}\label{optimal p'}
Let $p$ be a prime and let $u_1, u_2,\ldots, u_s$ be integers such that $u_s\geq u_{s-1}\geq \ldots \geq u_1$. Then there exists a $(py, p, \{K_0,\ldots, K_{p^{u_1-1}-1}\} , 1)$-BNCRDP of
size $\sum\limits_{i=1}^s p^{u_i-1}p^{u_{i+1}}\ldots p^{u_s}$ such that all elements of base blocks of each CRDP, together with $0$, form a complete system of representatives for the cosets of $y\mathbb{Z}_{py}$ in $\mathbb{Z}_{py}$ where $y=\prod\limits_{i=1}^s(p^{u_i}-1)$ and $K_i\subset \{1,2,\ldots,p\}$ for each $i$, $0\leq i< p^{u_1-1}$.
\end{corollary}

\begin{IEEEproof}
We shall prove that there exists such a BNCRDP by induction on $s$.

For $s=1$, the assertion holds by Corollary \ref{optimal p 3}. Assume that the assertion holds for $s=r$ and consider $s=r+1$. Let $x=\prod\limits_{i=1}^r(p^{u_i}-1)$ and $z=\sum\limits_{i=1}^r p^{u_i-1}p^{u_{i+1}}\ldots p^{u_r}$, assume that there exists a $(px, p, \{K'_0,\ldots, K'_{p^{u_1-1}-1}\}, 1)$-BNCRDP of size $z$ such that all elements of base blocks of each CRDP, together with $0$, form a complete system of representatives for the cosets of $x\mathbb{Z}_{px}$ in $\mathbb{Z}_{px}$ where $K'_i\subset \{1,2,\ldots,p\}$ for each $i$, $0\leq i< p^{u_1-1}$. Since $\max \limits_{0\leq k <z-1}\{\sum \limits_{j=0}^{p^{u_1-1}-1} |B_k^j|\}\leq p^{u_1} \leq p^{u_{r+1}}$ and $gcd(p^{u_{r+1}}-1, p)=1$, applying Theorem \ref{q-1 1} with $g=p$ and $s=1$ yields a $(px(p^{u_{r+1}}-1), p(p^{u_{r+1}}-1), \{K''_0,\ldots, K''_{p^{u_1-1}-1}\}, 1)$-BNCRDP of size $z p^{u_{r+1}}$ such that all elements of base blocks of each CRDP, together with $0,x,2x,\ldots, (p^{u_{r+1}}-2)x$, form a complete system of representatives for the cosets of $x(p^{u_{r+1}}-1) \mathbb{Z}_{px(p^{u_{r+1}}-1)}$ in $\mathbb{Z}_{px(p^{u_{r+1}}-1)}$ where $K''_i\subset \{1,2,\ldots,p\}$ for each $i$, $0\leq i< p^{u_1-1}$. By Corollary \ref{optimal p 3} and Lemma \ref{CRDPs=>CDPs 21}, we obtain a $(px(p^{u_{r+1}}-1), p, \{K_0,\ldots, K_{p^{u_1-1}-1}\}, 1)$-BNCRDP of size $zp^{u_{r+1}}+p^{u_{r+1}-1}$ such that all elements of base blocks of each CRDP, together with $0$, form a complete system of representatives for the cosets of $x(p^{u_{r+1}}-1) \mathbb{Z}_{px(p^{u_{r+1}}-1)}$ in $\mathbb{Z}_{px(p^{u_{r+1}}-1)}$ where $K_i=K'_i\cup K''_i\subset \{1,2,\ldots,p\}$ for each $i$, $0\leq i< p^{u_1-1}$. So, the conclusion holds by induction.
\end{IEEEproof}

Combing Corollary \ref{optimal p'} with Lemma \ref{CRDPs=>CDPs}, we obtain the following corollary.

\begin{corollary}\label{optimal p 4}
Let $p$ be a prime and let $u_1, u_2,\ldots, u_s$ be integers such that $u_s\geq u_{s-1}\geq \ldots \geq u_1$. If $p^{u_2}\geq p^{3u_1+1}u_2$, $s\geq 3$, $p^{u_2}\geq 2^s$ and $p^{u_1-1}>5$, then there exists a strictly optimal $(p\prod\limits_{i=1}^s(p^{u_i}-1),p^{u_1-1}, p; 1+\sum\limits_{i=1}^s p^{u_i}p^{u_{i+1}}\ldots p^{u_s})$-FHS set with respect to the Peng-Fan bounds.
\end{corollary}

\begin{IEEEproof}
Let ${\cal B}_i=\{\{0,1,\ldots,p-1\} \}$ for each $i$, $0\leq i<p^{m-1}$. Clearly,
${\cal B}=\{{\cal B}_i:~0\leq i<p^{m-1}\}$ is a $(p, \{\{p\},\ldots, \{p\}\},p)$-BNCDP of size $1$ such that ${\cal B}$ such that $d_a^{\cal B}\geq a$ for $1\leq a\leq p$. By Corollary \ref{optimal p'}, there exists a $(py, p, \{K'_0,\ldots, K'_{p^{u_1-1}-1}\}, 1)$-BNCRDP of size $\sum\limits_{i=1}^s p^{u_i-1}p^{u_{i+1}}\ldots p^{u_s}$ such that all elements of base blocks of each CRDP, together with $0$, form a complete system of representatives for the cosets of $y\mathbb{Z}_{py}$ in $\mathbb{Z}_{py}$ where $y=\prod\limits_{i=1}^s(p^{u_i}-1)$ and $K'_i\subset \{1,2,\ldots,p\}$ for each $i$, $0\leq i< p^{u_1-1}$. Applying  Lemma \ref{CRDPs=>CDPs} with ${\cal B}$, there exists a $(p\prod\limits_{i=1}^s(p^{u_i}-1),\{K_0,\ldots, K_{p^{u_1-1}-1}\}, p)$-BNCDP of size $1+\sum\limits_{i=1}^s p^{u_i}p^{u_{i+1}}\ldots p^{u_s}$, ${\cal D}$ such that $d_a^{\cal D}\geq a\prod\limits_{i=1}^s(p^{u_i}-1)$ for $1\leq a\leq p$.

Finally, we show $\left\lceil\frac{2InM-(I+1)Il}{(nM-1)M}\right\rceil=p$.

Since $s\geq 3$ and $u_1\leq u_2$, we have
{ \begin{equation}
\label{Bound p}
\begin{aligned}
&(p^{u_1}+1)p^{u_{2}}\ldots p^{u_s}\
<1+\sum\limits_{i=1}^s p^{u_i}p^{u_{i+1}}\ldots p^{u_s}\\
&< (p^{u_1}p^{u_2}+p^{u_2}+2)p^{u_{3}}\ldots p^{u_s}\ \leq (p^{u_1}+1+\frac{2}{p^{u_1}})p^{u_{2}}\ldots p^{u_s}.
\end{aligned}
\end{equation}
}
Since $u_2\leq u_3\leq \ldots \leq u_s$ and $p^{u_2}\geq 2^s$, we have
{ \begin{equation}
\label{p^m 1}
\begin{aligned}
&\prod\limits_{i=2}^s\frac{p^{u_i}}{p^{u_i}-1}<(1+\frac{1}{p^{u_2}-1})^{s-1} \ < 1+\frac{s-1}{p^{u_2}-1}+\frac{2^{s-1}}{(p^{u_2}-1)^2} \ < 1+\frac{s}{p^{u_2}-1}.
\end{aligned}
\end{equation}
}

Since $p^{u_2}\geq 2^s$, we have $s<pu_2$. Since $p^{u_2}\geq p^{3u_1+1}u_2$, it holds that
{ \small \begin{equation}
\label{p^m 2}
\begin{aligned}
\vspace{0.2cm}\frac{p^{u_2}-1}{s}\geq  \frac{ p^{3u_1+1}u_2-1}{s}>  \frac{ p^{3u_1}s-1}{s}>  p^{3u_1}-p^{2u_1}-4.
\end{aligned}
\end{equation}
}
Then,
{\small \begin{equation}
\label{Bound p2}
\begin{aligned}
&\prod\limits_{i=2}^s\frac{p^{u_i}}{p^{u_i}-1}<1+\frac{s}{p^{u_2}-1}\  \leq 1+ \frac{4}{p^{3u_1}-p^{2u_1}-4}\ =\frac{p^{3u_1}-p^{2u_1}}{p^{3u_1}-p^{2u_1}-4} \ =\frac{p^{2u_1}-p^{u_1}}{(p^{u_1}+1+\frac{2}{p^{u_1}})(p^{u_1}-2)}.
\end{aligned}
\end{equation}
}
In view of inequality (\ref{Bound p}) and inequality (\ref{Bound p2}), we have
\[
\begin{array}{l}
\vspace{0.2cm}p^{u_1}-2<\frac{p^{u_1}(p^{u_1}-1)}{(p^{u_1}+1+\frac{2}{p^{u_1}})}\times \prod\limits_{i=2}^s\frac{p^{u_i}-1}{p^{u_{i}}}\ <\frac{(p\prod\limits_{i=1}^s(p^{u_i}-1))p^{u_1-1}}{1+\sum\limits_{i=1}^s p^{u_i}p^{u_{i+1}}\ldots p^{u_s}}=\frac{nM}{l}\
<\frac{(p\prod\limits_{i=1}^s(p^{u_i}-1))p^{u_1-1}}{(p^{u_1}+1)p^{u_{2}}\ldots p^{u_s}}<p^{u_1}-1.
\end{array}
\]

By definition, we have
\[
\begin{array}{l}
\vspace{0.2cm}I=\left\lfloor \frac{nM}{l} \right \rfloor=\left\lfloor \frac{(p\prod\limits_{i=1}^s(p^{u_i}-1))p^{u_1-1}}{1+\sum\limits_{i=1}^s p^{u_i}p^{u_{i+1}}\ldots p^{u_s}} \right \rfloor \
 =p^{u_1}-2.
\end{array}
\]

It only needs to show that $\frac{2InM-(I+1)Il}{(nM-1)M}>p-1.$

Since $p^{u_1-1}\geq 5$, we have that
\[
\begin{array}{l}
\vspace{0.2cm}p^{3u_1-1}-2p^{2u_1-1}>p^{3u_1-1}-2p^{2u_1}\geq 3 p^{2u_1},\\ {\rm and}\ \
\frac{2}{p^{u_1}-2}<\frac{p^{2u_1-1}-p^{u_1}-2}{p^{2u_1}-3p^{u_1}+2}.
\end{array}
\]

In view of inequality (\ref{Bound p}), inequality (\ref{Bound p2}) and the last inequality, we have that
\[
\begin{array}{l}
\vspace{0.2cm}\frac{1+\sum\limits_{i=1}^s p^{u_i}p^{u_{i+1}}\ldots p^{u_s}}{\prod\limits_{i=1}^s(p^{u_i}-1)}\
< \frac{p^{u_1}+1+\frac{2}{p^{u_1}}}{p^{u_1}-1} \times \prod\limits_{i=2}^s\frac{p^{u_i}}{p^{u_i}-1}\
 <\frac{p^{u_1}+1+\frac{2}{p^{u_1}}}{p^{u_1}-1} \times \frac{p^{2u_1}-p^{u_1}}{(p^{u_1}+1+\frac{2}{p^{u_1}})(p^{u_1}-2)}\\
 = \frac{p^{u_1}}{p^{u_1}-2}= 1+\frac{2}{p^{u_1}-2}\
 \leq 1+ \frac{p^{2u_1-1}-p^{u_1}-2}{p^{2u_1}-3p^{u_1}+2}\
=\frac{p^{2u_1}-4p^{u_1}+p^{2u_1-1}}{p^{2u_1}-3p^{u_1}+2},
\end{array}
\]
and
{ \footnotesize
\[
\begin{array}{l}
\vspace{0.2cm} (p^{2u_1}-3p^{u_1}+2)(1+\sum\limits_{i=1}^s p^{u_i}p^{u_{i+1}}\ldots p^{u_s})
< (p^{2u_1}-4p^{u_1}+p^{2u_1-1})(\prod\limits_{i=1}^s(p^{u_i}-1)).
\end{array}
\]
}
Then, it holds that
\[
\begin{array}{l}
\vspace{0.2cm}\frac{2InM-(I+1)Il}{(nM-1)M}=\frac{I(2nM-(I+1)l)}{(nM-1)M}= \frac{(p^{u_1}-2)(2(p\prod\limits_{i=1}^s(p^{u_i}-1))p^{u_1-1}-(p^{u_1}-1)(1+\sum\limits_{i=1}^s p^{u_i}p^{u_{i+1}}\ldots p^{u_s}))}{((p\prod\limits_{i=1}^s(p^{u_i}-1))p^{u_1-1}-1)p^{u_1-1}}
>p-1.
\end{array}
\]

By Theorem \ref{optimal sets character}, ${\cal D}$ is a strictly optimal $(p\prod\limits_{i=1}^s(p^{u_i}-1),p^{m_1-1}, p; 1+\sum\limits_{i=1}^s p^{u_i}p^{u_{i+1}}\ldots p^{u_s})$-FHS set with respect to the Peng-Fan bounds. This completes the proof.
\end{IEEEproof}

\begin{corollary}\label{optimal p 5}
Let $p$ be an odd prime and let $u_1, u_2,\ldots, u_s, m$ be integers such that $u_s\geq u_{s-1}\geq \ldots \geq u_1$. If $p^{u_2}\geq p^{3u_1+1}u_2$, $s\geq 3$, $p^{u_2}\geq 2^s$, $p^{u_1-1}>5$ and $2^{p^m}>p^{3u_1}-p^{2u_1}$, then there exists a strictly optimal $(p(2^{p^m}-1)(\prod\limits_{i=1}^s(p^{u_i}-1)),p^{u_1-1}, p; 2^{p^m}(1+\sum\limits_{i=1}^s p^{u_i}p^{u_{i+1}}\ldots p^{u_s})-1)$-FHS set with respect to the Peng-Fan bounds.
\end{corollary}

\begin{IEEEproof}
Firstly, we prove that there exists a partition-type $(p(2^{p^m}-1),\{K'_0,\ldots,K'_{f-1}\},p)$-BNCDP of size $2^{p^m}-1$, ${\cal A}$ such that $d_i^{{\cal A}}\geq (2^{p^m}-1)i$ for $1\leq i\leq p$, where $K'_0=\cdots=K'_{f-1}=\{p\}$ and $f=\frac{2^{p^m}-2}{p}$.

Let $r$ be any prime such that $r|2^{p^m}-1$. We will show $p|r-1$. By Fermat little theorem, we have $r|2^{r-1}-1$.
Since gcd$(2^{r-1}-1, 2^{p^m}-1)=2^{gcd(r-1, p^m)}-1$ and gcd$(r-1, p^m)=p^a$ for some nonnegative integer $a$, we have $r|$ gcd$(2^{r-1}-1, 2^{p^m}-1)=2^{p^a}-1$. Then, $a$ is a positive integer and $p|r-1$. Applying Corollary \ref{uv1} and Lemma \ref{CRDPs=>CDPs 2},
there exists a partition-type $(p(2^{p^m}-1),\{K_0',\ldots,K_{f-1}'\},p)$-BNCDP of size $2^{p^m}-1$, ${\cal A}$ such that $d_i^{\cal A}\geq (2^{p^m}-1)i$ for $1\leq i\leq p$, where $K'_0=\cdots=K'_{f-1}=\{p\}$.

Secondly, we prove that there exists a partition-type $((p(2^{p^u}-1)(\prod\limits_{i=1}^s(p^{u_i}-1)),\{K^1_0,\ldots,K^1_{p^{u_1-1}-1}\},p)$-BNCDP of size $2^{p^m}(1+\sum\limits_{i=1}^s p^{u_i}p^{u_{i+1}}\ldots p^{u_s})-1$, ${\cal D}$ such that $d_i^{{\cal D}}\geq (2^{p^m}-1)(\prod\limits_{i=1}^s(p^{u_i}-1))i$ for $1\leq i\leq p$, where $K^1_0=\cdots=K^1_{p^{u_1-1}-1}\subset \{1,2,\ldots p\}$ and $f=\frac{2^{p^m}-2}{p}$.

By Corollary \ref{optimal p'}, there exists a $(py, p, \{K_0,\ldots, K_{p^{u_1-1}-1}\} , 1)$-BNCRDP of
size $\sum\limits_{i=1}^s p^{u_i-1}p^{u_{i+1}}\ldots p^{u_s}$ such that all elements of base blocks of each CRDP, together with $0$, form a complete system of representatives for the cosets of $y\mathbb{Z}_{py}$ in $\mathbb{Z}_{py}$ where $y=\prod\limits_{i=1}^s(p^{u_i}-1)$ and $K_i\subset \{1,2,\ldots,p\}$ for each $i$, $0\leq i< p^{u_1-1}$. By Fermat little theorem, we have $2^p\equiv 2\mod p$ for any odd prime $p$. Then, it holds that $2^{p^m}\equiv 2^{p^{m-1}}\equiv \ldots \equiv 2\pmod p$ and $gcd(2^{p^m}-1,p)=1$. Since $\max \limits_{k}\{\sum \limits_{j=0}^{p^{u_1-1}-1} |B_k^j|\}\leq p^{u_1}\leq 2^{p^m}$, by Theorem \ref{q-1 1} with $g=p$ and $s=1$ yields a $(py(2^{p^m}-1), p(2^{p^m}-1), \{K''_0,\ldots, K''_{p^{u_1-1}-1}\} , 1)$-BNCRDP of size $2^{p^m}\sum\limits_{i=1}^s p^{u_i-1}p^{u_{i+1}}\ldots p^{u_s}$ such that all elements of base blocks of each CRDP, together with $0,y, \ldots, (2^{p^m}-2)y$, form a complete system of representatives for the cosets of $y(2^{p^m}-1)\mathbb{Z}_{py(2^{p^m}-1)}$ in $\mathbb{Z}_{py(2^{p^m}-1)}$. Since $2^{p^m}-2>p^{u_1}$, we have $f>p^{u_1-1}$. Since there exists a partition-type $(p(2^{p^m}-1),\{K'_0,\ldots,K'_{p^{u_1-1}-1}\},p)$-BNCDP, ${\cal A}$ such that $d_i^{{\cal A}}\geq (2^{p^m}-1)i$ for $1\leq i\leq p$, by Lemma \ref{CRDPs=>CDPs} with $g=p(2^{p^{m}}-1)$ and $s=2^{p^{m}}-1$ yields a partition-type $(p(2^{p^m}-1)(\prod\limits_{i=1}^s(p^{u_i}-1)),\{K^1_0,\ldots,K^1_{p^{u_1-1}-1}\},p)$-BNCDP of size $2^{p^m}(1+\sum\limits_{i=1}^s p^{u_i}p^{u_{i+1}}\ldots p^{u_s})-1$, ${\cal D}$ such that $d_i^{{\cal D}}\geq ((2^{p^m}-1)(\prod\limits_{i=1}^s(p^{u_i}-1)))i$ for $1\leq i\leq p$, where $K^1_i=K'_i\cup K''_i\subset \{1,2,\ldots p\}$.

Finally, we show $\left\lceil\frac{2InM-(I+1)Il}{(nM-1)M}\right\rceil=p$.

In view of inequality (\ref{p^m 1}), inequality (\ref{p^m 2}), we have
\[
\begin{array}{l}
\vspace{0.2cm}\prod\limits_{i=2}^s \frac{p^{u_i}}{p^{u_i}-1}< 1+\frac{s}{p^{u_2}-1}, \ {\rm and}\ \frac{p^{u_2}-1}{s} > p^{3u_1}-p^{2u_1}-4.
\end{array}
\]
Since $2^{p^m}>p^{3u_1}-p^{2u_1}$, it holds that
{\small \begin{equation}
\label{Bound p4}
\begin{aligned}
&\prod\limits_{i=2}^s \frac{p^{u_i}}{p^{u_i}-1}<1+\frac{s}{p^{u_2}-1}\ <1+\frac{1}{p^{3u_2}-p^{2u_2}-4}\\
& \leq 1+ \frac{4}{p^{3u_1}-p^{2u_1}-4}(1-\frac{1}{2^{p^m}})-\frac{1}{2^{p^m}}\
=\frac{(2^{p^m}-1)(p^{2u_1}-p^{u_1})}{2^{p^m}(p^{u_1}+1+\frac{2}{p^{u_1}})(p^{u_1}-2)}.
\end{aligned}
\end{equation}
}
In view of inequality (\ref{Bound p}) and inequality (\ref{Bound p4}), we have

\[
\begin{array}{l}
\vspace{0.2cm}p^{u_1}-2<\frac{(2^{p^m}-1)p^{u_1}(p^{u_1}-1)}{2^{p^m}(p^{u_1}+1+\frac{2}{p^{u_1}})}\times \prod\limits_{i=2}^s\frac{p^{u_i}-1}{p^{u_{i}}}\\
\vspace{0.2cm}<\frac{(2^{p^m}-1)(p\prod\limits_{i=1}^s(p^{u_i}-1))p^{u_1-1}}{2^{p^m}(1+\sum\limits_{i=1}^s p^{u_i}p^{u_{i+1}}\ldots p^{u_s})-1}=\frac{nM}{l}\\
<\frac{(2^{p^m}-1)(p\prod\limits_{i=1}^s(p^{u_i}-1))p^{u_1-1}}{2^{p^m}(p^{u_1}+1)p^{u_{2}}\ldots p^{u_s}}<p^{u_1}-1.
\end{array}
\]

By definition, we have
\[
\begin{array}{l}
\vspace{0.2cm}I=\left\lfloor \frac{nM}{l} \right \rfloor=\left\lfloor \frac{(2^{p^m}-1)(p\prod\limits_{i=1}^s(p^{u_i}-1))p^{u_1-1}}{2^{p^m}(1+\sum\limits_{i=1}^s p^{u_i}p^{u_{i+1}}\ldots p^{u_s})-1} \right \rfloor
 =p^{u_1}-2.
\end{array}
\]

It only needs to show that $\frac{2InM-(I+1)Il}{(nM-1)M}>p-1.$

Since $p^{u_1-1}\geq 5$, we have that $$p^{3u_1-1}-2p^{2u_1-1}>p^{3u_1-1}-2p^{2u_1}>3 p^{2u}.$$

In view of inequality (\ref{Bound p}), inequality (\ref{Bound p4}) and the last inequality, we have

\[
\begin{array}{l}
\vspace{0.2cm}\frac{2^{p^m}(1+\sum\limits_{i=1}^s p^{u_i}p^{u_{i+1}}\ldots p^{u_s})-1}{(2^{p^m}-1)(\prod\limits_{i=1}^s(p^{u_i}-1))}\\
\vspace{0.2cm}< \frac{2^{p^m}(p^{u_1}+1+\frac{2}{p^{u_1}})}{(2^{p^m}-1)(p^{u_1}-1)} \times \prod\limits_{i=2}^s\frac{p^{u_i}}{p^{u_i}-1}\\
\vspace{0.2cm} <\frac{2^{p^m}(p^{u_1}+1+\frac{2}{p^{u_1}})}{(2^{p^m}-1)(p^{u_1}-1)} \times \frac{(2^{p^m}-1)(p^{2u_1}-p^{u_1})}{2^{p^m}(p^{u_1}+1+\frac{2}{p^{u_1}})(p^{u_1}-2)}\\
\vspace{0.2cm} = \frac{p^{u_1}}{p^{u_1}-2}=1+\frac{2}{p^{u_1}-2}\\
\vspace{0.2cm} \leq1+ \frac{p^{2u_1-1}-p^{u_1}-2}{p^{2u_1}-3p^{u_1}+2}\\
 =\frac{p^{2u_1}-4p^{u_1}+p^{2u_1-1}}{p^{2u_1}-3p^{u_1}+2},
\end{array}
\]
and
{ \footnotesize
\[
\begin{array}{l}
(p^{2u_1}-3p^{u_1}+2) (2^{p^m}(1+\sum\limits_{i=1}^s p^{u_i}p^{u_{i+1}}\ldots p^{u_s})-1)
< (p^{2u_1}-4p^{u_1}+p^{2u_1-1})((2^{p^m}-1)(\prod\limits_{i=1}^s(p^{u_i}-1))).
\end{array}
\]
}
Then, it holds that
{
\[
\begin{array}{l}
\vspace{0.2cm}\frac{2InM-(I+1)Il}{(nM-1)M}=\frac{2InM}{(nM-1)M}-\frac{(I+1)Il}{(nM-1)M}=\\ \frac{2(p^{u_1}-2)(2^{p^m}-1)(p\prod\limits_{i=1}^s(p^{u_i}-1))p^{u_1-1}}{((2^{p^m}-1)(p\prod\limits_{i=1}^s(p^{u_i}-1))p^{u_1-1}-1)p^{u_1-1}}
-\frac{(p^{u_1}-2)(p^{u_1}-1)(2^{p^m}(1+\sum\limits_{i=1}^s p^{u_i}p^{u_{i+1}}\ldots p^{u_s})-1)}{((2^{p^m}-1)(p\prod\limits_{i=1}^s(p^{u_i}-1))p^{u_1-1}-1)p^{u_1-1}}\\
>p-1.
\end{array}
\]
}
By Theorem \ref{optimal sets character}, ${\cal D}$ is a strictly optimal $(p(2^{p^m}-1)(\prod\limits_{i=1}^s(p^{u_i}-1)),p^{u_1-1}, p; 2^{p^m}(1+\sum\limits_{i=1}^s p^{u_i}p^{u_{i+1}}\ldots p^{u_s})-1)$-FHS set with respect to the Peng-Fan bounds. This completes the proof.

\end{IEEEproof}

\begin{corollary}\label{euv}
Let $v$ be an odd integer of the form $v=p_1^{m_1}p_2^{m_2}\cdots p_s^{m_s}$ for $s$ positive integers $m_1,m_2,\ldots,m_s$ and $s$ primes $p_1,p_2,\ldots,p_s$ such that $p_1<p_2<\cdots<p_s$, and let $p$ be a prime such that $p|p_i-1$ for each $i$, $0<i\leq s$. Set $f=\frac{p_1-1}{p}$. Let $m$ be an integer such that $p^m >p_1-1$. If $p^m-2(1+p)>0$ and $f\geq 2$, then there exists a strictly optimal $(pv(p^m-1),f, p; vp^m)$-FHS set with respect to the Peng-Fan bounds.
\end{corollary}

\begin{IEEEproof}
Firstly, we prove that there exists a partition-type $(pv(p^m-1),\{K'_0,\ldots,K'_{f-1}\},p)$-BNCDP of size $vp^m$, ${\cal D}$ such that $d_i^{{\cal D}}\geq v(p^m-1)i$ for $1\leq i\leq p$, where $K'_0=\cdots=K'_{f-1}=\{ p,p-1\}$.

By Lemma \ref{cyclotomic construction} with $u=e=p$, there exists a $(pv,p,\{K_0,\ldots,K_{f-1}\},1)$-BNCRDP of size
$\frac{v-1}{p}$ such that all elements of base blocks of each CRDP, together with $0$, form a complete system of representatives for the cosets of $v\mathbb{Z}_{pv}$ in $\mathbb{Z}_{pv}$ where $K_0=\cdots=K_{f-1}=\{p\}$. Since $\max \limits_{0\leq k <\frac{v-1}{p}}\{\sum \limits_{j=0}^{f-1} |B_k^j|\}= p_1-1< p^m$ and gcd$(p^m-1,p)=1$, applying Theorem \ref{q-1 1} with $g=p$ and $s=1$ yields a $(pv(p^m-1),p(p^m-1),\{K'_0,K'_1,\ldots,K'_{f-1}\},1)$-BNCRDP of size $(v-1)p^{m-1}$ such that all elements of base blocks of each CRDP, together with $0,v,\ldots,(p^m-2)v$, form a complete system of representatives for the cosets of $v(p^m-1)\mathbb{Z}_{pv(p^m-1)}$ in $\mathbb{Z}_{pv(p^m-1)}$ where $K'_0=K'_1=\ldots=K'_{f-1}=\{p-1,p\}$. By Corollary \ref{optimal p 1} and Theorem \ref{optimal sets character}, there exists a partition-type $(p(p^m-1),\{K'_0,\ldots,K'_{f-1}\},p)$-BNCDP of size $p^m$, ${\cal A}$ with $d_i^{{\cal A}}\geq (p^m-1)i$ for $1\leq i\leq p$ where $K'_0=\cdots=K'_{p^{m-1}}=\{p, p-1\}$. Since $p^m> p_1-1$, we have $p^{m-1}\geq f$. By applying Lemma \ref{CRDPs=>CDPs} with $g=p(p^m-1)$ and $s=p^m-1$ we obtain a partition-type $(pv(p^m-1),\{K'_0,\ldots,K'_{f-1}\},p)$-BNCDP of size $vp^m$, ${\cal D}$ with $d_i^{{\cal D}}\geq (p^m-1)vi$ for $1\leq i\leq p$, where $K'_0=\cdots=K'_{f-1}=\{p-1,p\}$.

Finally, we show $\left\lceil\frac{2InM-(I+1)Il}{(nM-1)M}\right\rceil=p$.

Since $p^m> p_1-1=pf$, we have
\[
\begin{array}{l}
\vspace{0.2cm}I=\left\lfloor \frac{nM}{l} \right \rfloor=\left\lfloor \frac{pv(p^m-1)f}{vp^m} \right \rfloor  =\left\lfloor pf-\frac{pf}{p^m} \right \rfloor  = pf-1.
\end{array}
\]

By definition, we have
{\small \begin{equation}
\label{Bound e1}
\begin{aligned}
\left\lceil\frac{2InM-(I+1)Il}{(nM-1)M}\right\rceil &=\left\lceil\frac{2(pf-1)pv(p^m-1)f-pf(pf-1)vp^m}{(pv(p^m-1)f-1)f}\right\rceil \\
&= \left\lceil p- \frac{p(pfv-1+vp^m-2v)}{pv(p^m-1)f-1}\right\rceil.
\end{aligned}
\end{equation}
}
Since $p^m-2(1+p)\geq 0$ and $f\geq 2$, we have
\[
\begin{array}{l}
\vspace{0.2cm}p^mf-f-pf-p^m+2 =(p^m-(1+p))(f-1)+1-p
> (1+p)+1-p>0.
\end{array}
\]
Then,
\[
\begin{array}{l}
0< \frac{p(pfv-1+vp^m-2v)}{pv(p^m-1)f-1}<1.
\end{array}
\]
It follows from (\ref{Bound e1}) that $\left\lceil\frac{2InM-(I+1)Il}{(nM-1)M}\right\rceil=p$. By Theorem \ref{optimal sets character}, ${\cal D}$ is a strictly optimal $(pv(p^m-1),f, p; vp^m)$-FHS set with respect to the Peng-Fan bounds. This completes the proof.
\end{IEEEproof}

When we replace the BNCRDP in Theorem \ref{q-1 1} with a $(g,\{K_0,K_1,\ldots,K_{M-1}\},1)$-BNCDP such that all elements of base blocks of each CDP form a complete system of representatives for the cosets of $s\mathbb{Z}_{g}$ in $\mathbb{Z}_{g}$ where $s|g$, the same procedure yields a new partition-type BNCDP. Since the proof is similar to that of Theorem \ref{q-1 1}, we omit it here.

\begin{theorem}\label{q-1 2}
Assume that $\{{\cal B}_0,\ldots,{\cal B}_{M-1}\}$ is a $(g,\{K_0,K_1,\ldots,K_{M-1}\},1)$-BNCDP of size $u$ such that all elements of base blocks of ${\cal B}_j$ form a complete system of representatives for the cosets of $s\mathbb{Z}_{g}$ in $\mathbb{Z}_{g}$for $0\leq j <M$, where $s|g$ and ${\cal B}_j=\{B_0^j,B_1^j,\ldots,B_{u-1}^j\}$. Let $q$ be a prime power such that $q\geq\max \limits_{0\leq k <u}\{\sum \limits_{j=0}^{M-1} |B_k^j|\}$ and $gcd(q-1,\frac{g}{s})=1$, then there also exists a  $(g(q-1),\{K'_0,K'_1,\ldots,K'_{M-1}\},1)$-BNCDP of size $uq$, ${\cal B}'=\{{\cal B}'_0,\ldots,{\cal B}'_{M-1}\}$ such that all elements of base blocks of ${\cal B}_{j}'$ form a complete system of representatives for the cosets of $s(q-1)\mathbb{Z}_{g(q-1)}$ in $\mathbb{Z}_{g(q-1)}$ for $0\leq j <M$, where $K'_i\subset K_i\cup \{k-1:~ k\in K_i\}$ for each $i$,\ $0\leq i<M$.
\end{theorem}

Combining Theorem \ref{q-1 2}, Lemma \ref{CRDPs=>CDPs 2} and Corollary \ref{d2} together, we obtain the following corollary.

\begin{corollary}\label{d4}
Let $q$ be a prime power, $d, m$ positive integers such that $m\geq 4$, $d|q-1$ and gcd$(m,d)=1$. Let $q'$ be a prime power such that $q'\geq q$ and $gcd(q'-1,\frac{q-1}{d})=1$. If $d\geq 2$ and $q'>q+1$, then there exists a strictly optimal $(\frac{(q'-1)(q^m-1)}{d}, d, \frac{q-1}{d}; (q^{m-1}-1+\frac{q-1}{d})q')$-FHS set with respect to the Peng-Fan bounds.
\end{corollary}

\begin{IEEEproof}
By using Corollary \ref{d2}, there exists a $(\frac{q^m-1}{d}, \{K'_0,\ldots, K'_{d-1}\}, 1)$-BNCDP of size $\frac{(q^{m-1}-1)d}{q-1}+1$ such that all elements of base blocks of each CDP form a complete system of representatives for the cosets of $\frac{q^{m}-1}{q-1} \mathbb{Z}_{\frac{q^{m}-1}{d}}$ in $\mathbb{Z}_{\frac{q^{m}-1}{d}}$.  In view of equation (\ref{sum}), we have $\max \limits_{0\leq k <\frac{d(q^{m-1}-1)}{q-1}+1}\{\sum \limits_{j=0}^{d-1} |B_k^j|\}= q$. Since $q'$ be a prime power such that $q' \geq q= \max \limits_{0\leq k < \frac{d(q^{m-1}-1)}{q-1}+1}\{\sum \limits_{j=0}^{d-1} |B_k^j|\}$ and $gcd(q'-1,\frac{q-1}{d})=1$, by Theorem \ref{q-1 2} with $g=\frac{q^m-1}{d}$ and $s=\frac{q^m-1}{q-1}$ yields a $(\frac{(q'-1)(q^{m}-1)}{d},\{K_0,\ldots,K_{d-1}\},1 )$-BNCDP of size $(\frac{(q^{m-1}-1)d}{q-1}+1)q'$ such that all elements of base blocks of each CDP form a complete system of representatives for the cosets of $\frac{(q'-1)(q^{m}-1)}{q-1} \mathbb{Z}_{\frac{(q'-1)(q^{m}-1)}{d}}$ in $\mathbb{Z}_{\frac{(q'-1)(q^{m}-1)}{d}}$. By applying Lemma \ref{CRDPs=>CDPs 2}, we obtain a partition-type $(\frac{(q'-1)(q^{m}-1)}{d},\{K_0,\ldots,K_{d-1}\},\frac{q-1}{d})$-BNCDP of size $(q^{m-1}-1+\frac{q-1}{d})q'$, ${\cal D}$ such that $d_i^{\cal D} \geq i\frac{(q'-1)(q^{m}-1)}{q-1}$ for $1\leq i\leq \frac{q-1}{d}$.

Finally, we show $\left\lceil\frac{2InM-(I+1)Il}{(nM-1)M}\right\rceil=\frac{q-1}{d}$.

Since $d\geq2$, $m\geq 4$ and $q'\geq q+2$, we have
\[
\begin{array}{l}
\vspace{0.2cm}q' ((q^m-1)-(q-1)(q^{m-1}-1+\frac{q-1}{d}))\\
\vspace{0.2cm}=q'(q^{m-1}+q+\frac{q-1}{d}-\frac{q(q-1)}{d}-2)\\
\vspace{0.2cm}\geq q'(q-1)q^{m-2} \geq (q+2)(q-1)q^{m-2} \\
\geq q^{m}-1.
\end{array}
\]
Then, it holds that
\[
\begin{array}{l}
\vspace{0.2cm}I=\left\lfloor \frac{nM}{l} \right \rfloor=\left\lfloor \frac{\frac{(q'-1)(q^{m}-1)}{d} \cdot d}{(q^{m-1}-1+\frac{q-1}{d})q'} \right \rfloor  =q-1.
\end{array}
\]

Since $m\geq 4$ and $q'\geq q+2$, we have

\[
\begin{array}{l}
\vspace{0.2cm}q' (d(q^m-1)-(q-1)(1+\frac{(q-1)q}{d}-q))\\
\vspace{0.2cm}\geq q'(d(q^m-1)-(q-1)^3)
\geq dq'(q-1)q^{m-1}  \geq d(q+2)(q-1)q^{m-1} \\
\geq dq^{m}+(q-1)(q^m-2).
\end{array}
\]

Then, it holds that
\[
\begin{array}{l}
\vspace{0.3cm} 0<\frac{(q-1)(q'+\frac{q-1}{d}qq'-qq'+q^m-2)}{((q'-1)(q^{m}-1)-1)d}<1,\  {\rm and} \\
\vspace{0.3cm}\left\lceil\frac{2InM-(I+1)Il}{(nM-1)M}\right\rceil =\left\lceil\frac{2(q-1)\frac{(q'-1)(q^{m}-1)}{d} \cdot d-q(q-1)(q^{m-1}-1+\frac{q-1}{d})q'}{(\frac{(q'-1)(q^{m}-1)}{d} \cdot d-1)d}\right\rceil \\
\vspace{0.3cm}\hspace{2.5cm}= \left\lceil \frac{q-1}{d}-\frac{(q-1)(q'+\frac{q-1}{d}qq'-qq'+q^m-2)}{((q'-1)(q^{m}-1)-1)d}\right\rceil \\
 \hspace{2.5cm} = \frac{q-1}{d}.
\end{array}
\]

By Theorem \ref{optimal sets character}, ${\cal D}$ is a strictly optimal $(\frac{(q'-1)(q^m-1)}{d}, d, \frac{q-1}{d}; (q^{m-1}-1+\frac{q-1}{d})q')$-FHS set with respect to the Peng-Fan bounds. This completes the proof.
\end{IEEEproof}

Combining Theorem \ref{q-1 2}, Lemma \ref{CRDPs=>CDPs 2} and Theorem \ref{p(p^m-1) 2} together, we yield the following corollary.

\begin{corollary}\label{uv4}
Let $p$ be a prime and let $m$ be an integer with $m>1$.  Let $q$ be a prime power such that $q\geq p^m$ and $gcd(q-1,p)=1$. If $p^{m}-3p\geq 1$, then there exists a strictly optimal $((q-1)p(p^m-1),p^{m-1}, p; p^{m}q)$-FHS set with respect to the Peng-Fan bounds.
\end{corollary}

\begin{IEEEproof}
By using Theorem \ref{p(p^m-1) 2}, there exists a $(p(p^m-1),\{K'_0,\ldots,K'_{p^{m-1}-1}\},1 )$-BNCDP of size $p^{m-1}$ such that all elements of base blocks of each CDP form a complete system of representatives for the cosets of $(p^m-1) \mathbb{Z}_{p(p^m-1)}$ in $\mathbb{Z}_{p(p^m-1)}$ where $K'_0=\cdots=K'_{p^{m-1}-1}=\{p,p-1\}$. Since $q$ be a prime power such that $q\geq p^m \geq \max \limits_{0\leq k <p^{m-1}}\{\sum \limits_{j=0}^{p^{m-1}-1} |B_k^j|\}$ and $gcd(q-1,p)=1$, by Theorem \ref{q-1 2} with $g=p(p^m-1)$ and $s=p^m-1$ yields a $(p(p^m-1)(q-1),\{K_0,\ldots,K_{p^{m-1}-1}\},1 )$-BNCDP of size $p^{m-1}q$ such that all elements of base blocks of each CDP form a complete system of representatives for the cosets of $(p^m-1)(q-1) \mathbb{Z}_{p(p^m-1)(q-1)}$ in $\mathbb{Z}_{p(p^m-1)(q-1)}$ where $K_0=\cdots=K_{p^{m-1}-1}=\{p,p-1,p-2\}$. By applying Lemma \ref{CRDPs=>CDPs 2}, we obtain a partition-type $(p(p^m-1)(q-1),\{K_0,\ldots,K_{p^{m-1}-1}\},p)$-BNCDP of size $p^mq$, ${\cal D}$ such that $d_i^{\cal D} \geq i(p^m-1)(q-1)$ for $1\leq i\leq p$.

Finally, we show $\left\lceil\frac{2InM-(I+1)Il}{(nM-1)M}\right\rceil=p$.

Since $q\geq p^m$, we have
\[
\begin{array}{l}
\vspace{0.2cm}I=\left\lfloor \frac{nM}{l} \right \rfloor=\left\lfloor \frac{p(p^m-1)(q-1)\cdot p^{m-1}}{p^{m}q} \right \rfloor  =p^m-2.
\end{array}
\]

Since $p^{m}\geq 3p+1$ and $q\geq p^m$, we have
\[
\begin{array}{l}
\vspace{0.2cm}q(p^m-2p)-p^m(p+1)+4p>0,\ {\rm and} \\
0<\frac{p((p^m-1)(p^m+2q-4)-1)}{(q-1)(p^m-1) p^{m}-1}<1.
\end{array}
\]

Then,
\[
\begin{array}{l}
\vspace{0.2cm}\left\lceil\frac{2InM-(I+1)Il}{(nM-1)M}\right\rceil
=\left\lceil\frac{2(p^m-2)(q-1)p(p^m-1) \cdot p^{m-1}-(p^m-2)(p^m-1)p^{m}q}{((q-1)p(p^m-1) \cdot p^{m-1}-1)p^{m-1}}\right\rceil \\
\vspace{0.2cm}\hspace{2.5cm}= \left\lceil p-\frac{p((p^m-1)(p^m+2q-4)-1)}{(q-1)(p^m-1) p^{m}-1}\right\rceil \\
\hspace{2.5cm}=p.
\end{array}
\]

By Theorem \ref{optimal sets character}, ${\cal D}$ is a strictly optimal $((q-1)p(p^m-1),p^{m-1}, p; p^{m}q)$-FHS set with respect to the Peng-Fan bounds. This completes the proof.
\end{IEEEproof}

Combining Theorem \ref{q-1 2} and  Lemma \ref{CRDPs=>CDPs 2} with Corollary \ref{uv1} give the following corollary.

\begin{corollary}\label{uv3}
Let $v$ be a positive integer of the form $v=p_1^{m_1}p_2^{m_2}\cdots p_s^{m_s}$ for $s$ positive integers $m_1,m_2,\ldots,m_s$ and $s$ distinct primes $ p_1,p_2,\ldots,p_s$ with $p_1<p_2<\cdots<p_s$. Let $e$ be a common factor of $p_1-1, p_2-1, \ldots, p_s-1$ and $e>1$, and let $f=\min \{\frac{p_i-1}{e}\colon 1\leq i\leq s\}$. Let $q$ be a prime power such that $q> p_1-1$ and $gcd(q-1,e)=1$. If $v\geq e^3f^2$, $q\geq 2e+5$ and $f>1$, then there exists a strictly optimal $(ev(q-1),f, e; (v-1+e)q)$-FHS set with respect to the Peng-Fan bounds.
\end{corollary}

\begin{IEEEproof}
By Corollary \ref{uv1} with $u=e$, there exists an $(ev,\{K'_0,\ldots,K'_{f-1}\},1)$-BNCDP of size $\frac{v-1}{e}+1$ such that all elements of base blocks of each CDP form a complete system of representatives for the cosets of $v \mathbb{Z}_{ev}$ in $\mathbb{Z}_{ev}$ where $K'_0=\cdots=K'_{f-1}=\{e,1\}$.  Since $q$ be a prime power such that $q> p_1-1 \geq \max \limits_{0\leq k <\frac{v-1}{e}+1}\{\sum \limits_{j=0}^{f-1} |B_k^j|\}$ and $gcd(q-1,e)=1$, by Theorem \ref{q-1 2} with $g=ve$ and $s=v$ yields an $(ev(q-1),\{K_0,\ldots,K_{f-1}\},1 )$-BNCDP of size $(\frac{v-1}{e}+1)q$ such that all elements of base blocks of each CDP form a complete system of representatives for the cosets of $v(q-1) \mathbb{Z}_{ev(q-1)}$ in $\mathbb{Z}_{ev(q-1)}$ where $K_0=\cdots=K_{f-1}=\{e,e-1,1\}$. By applying Lemma \ref{CRDPs=>CDPs 2}, we obtain a partition-type $(ev(q-1),\{K_0,\ldots,K_{f-1}\},e)$-BNCDP of size $(v-1+e)q$, ${\cal D}$ such that $d_i^{\cal D} \geq iv(q-1)$ for $1\leq i\leq e$.

Finally, we show $\left\lceil\frac{2InM-(I+1)Il}{(nM-1)M}\right\rceil=e$.

Since $q>p_1-1=ef$ and $v\geq e^3f^2$, we have
\[
\begin{array}{l}
\vspace{0.2cm}q(v-1+e-e^2f+ef)\\
\vspace{0.2cm}=q(v-(e-1)(ef-1)) \geq (ef+1)(v-(e-1)(ef-1))\\
\vspace{0.2cm}=efv+v-(ef+1)(ef-1)(e-1) =efv+v-(e^2f^2-1)(e-1)\\
\vspace{0.2cm}>efv.
\end{array}
\]
Then, it holds that
\[
\begin{array}{l}
\vspace{0.2cm}I=\left\lfloor \frac{nM}{l} \right \rfloor=\left\lfloor \frac{ev(q-1) \cdot f}{(v-1+e)q} \right \rfloor   =\left\lfloor ef- \frac{ef(v-q+eq)}{(v-1+e)q} \right \rfloor  =ef-1.
\end{array}
\]

Since $f\geq 2$, $v\geq e^3f^2$ and $q\geq 2e+5$, we have
\[
\begin{array}{l}
\vspace{0.2cm}v(q-1)f-(efv+e^2fq+vq)\\
\vspace{0.2cm}=fqv-e^2fq-efv-fv-qv\\
\vspace{0.2cm}= (f-1)qv- (ef+f)v- e^2fq\\
\vspace{0.2cm}\geq \frac{f}{2}(qv-2(e+1)v-2e^2q)\\
\vspace{0.2cm}=\frac{f}{2}((q-2(e+1))(v-2e^2)-4e^2(e+1))\\
\vspace{0.2cm}\geq \frac{f}{2}(3(4e^3-4e^2)-4e^2(e+1))\\
\vspace{0.2cm}=4fe^2(e-2)\\
\vspace{0.2cm}\geq 0.
\end{array}
\]
Then,
$$0<\frac{e(efv+e^2fq+vq-efq-2v-eq+q-1)}{ev(q-1) f-1}<1$$
Hence, it holds that
\[
\begin{array}{l}
\vspace{0.2cm}\left\lceil\frac{2InM-(I+1)Il}{(nM-1)M}\right\rceil =\left\lceil\frac{2(ef-1)ev(q-1) \cdot f-ef(ef-1)(v-1+e)q}{(ev(q-1) \cdot f-1)f}\right\rceil \\
\vspace{0.2cm}\hspace{2.5cm}= \left\lceil e-\frac{e(efv+e^2fq+vq-efq-2v-eq+q-1)}{ev(q-1) f-1}\right\rceil \\
\hspace{2.5cm} =e.
\end{array}
\]

By Theorem \ref{optimal sets character}, ${\cal D}$ is a strictly optimal $(ev(q-1),f, e; (v-1+e)q)$-FHS set with respect to the Peng-Fan bounds. This completes the proof.
\end{IEEEproof}

\section{Concluding Remarks}  %
\label{concl}                               %

In this paper, several BNCDPs and BNCRDPs such that both of them have a special property were presented by trace functions and discrete logarithm. Three recursive constructions of strictly optimal FHS sets were obtained. Combing these BNCDPs, BNCRDPs and three recursive constructions, we obtained infinitely many families of new FHS sets having strictly optimal Hamming correlation with respect to the Peng-Fan bounds. Finally, we must point out that it would be possible to obtain more strictly optimal FHS sets from our recursive constructions, provided that we can find new appropriate BNCDPs or BNCRDPs such that both of them have the special property.

\ifCLASSOPTIONcaptionsoff
  \newpage
\fi

\end{document}